\documentclass[lettersize,journal]{IEEEtran}

\makeatletter
\def\ps@headings{%
	\def\@oddhead{\mbox{}\scriptsize\rightmark \hfil \thepage}%
	\def\@evenhead{\scriptsize\thepage \hfil \leftmark\mbox{}}%
	\def\@oddfoot{}%
	\def\@evenfoot{}}
\makeatother
\usepackage{amsmath}

\usepackage{amssymb}
\usepackage{tabularx}
\usepackage[skip=1ex]{caption}
\usepackage{multirow}
\usepackage[style=plain]{floatrow}
\usepackage{rotating}
\usepackage{graphicx}
\usepackage{tabularx}
\usepackage{array}
\usepackage{mathrsfs}
\usepackage{amsthm} 
\usepackage{amsfonts}
\usepackage{multicol}
\usepackage{algpseudocode}
\usepackage{csquotes}
\usepackage{tabu}
\usepackage{float}
\usepackage{url}
\usepackage{blindtext}
\usepackage{cite}
\usepackage{lipsum}
\usepackage{mathtools}
\usepackage{cuted}
\usepackage{etoolbox}
\usepackage{flushend}
\usepackage{subcaption}
\usepackage{stackengine}

\usepackage[ruled,vlined,linesnumbered,resetcount]{algorithm2e}
\SetKwRepeat{Do}{do}{while}%
\usepackage{soul}
\usepackage{color}
\usepackage{hyperref}

\usepackage[T1]{fontenc}

\usepackage{tikz}
\usetikzlibrary{shapes,shapes.geometric,arrows,fit,calc,positioning,automata,shadows,trees}
\tikzstyle{b1} = [rectangle, draw, fill=white,
text width=13em, text centered, node distance=1.5cm, inner sep=0pt, minimum height=2em]
\tikzstyle{b2} = [rectangle, draw, fill=white,
text width=10em, text centered, node distance=1.5cm, inner sep=0pt, minimum height=2em]
\tikzstyle{b3} = [rectangle, draw, fill=white,
text width=9em, text centered, node distance=1.5cm, inner sep=1pt, minimum height=2em]
\tikzstyle{c1} = [diamond, draw, fill=white, text width=8em, text centered, node distance=3cm, inner sep=0pt,minimum height=2em]
\tikzstyle{line} = [draw, -latex']
\tikzset{
	basic/.style  = {draw, text width=2cm, drop shadow, font=\sffamily, rectangle},
	test/.style={base, diamond, aspect=2, align=center, inner sep=-1ex},
	root/.style   = {basic, rounded corners=2pt, thin, align=center,
		fill=green!30},
	level 2/.style = {basic, rounded corners=6pt, thin,align=center, fill=green!60,
		text width=8em},
	level 3/.style = {basic, thin, align=left, fill=pink!60, text width=6.5em}
}

\newtheorem{lem}{Lemma}
\newtheorem{assumption}{Assumption}

\newtheorem{thm}{Theorem}

\usetikzlibrary{shapes,arrows,positioning}
\makeatletter
\renewcommand{\SetKwInOut}[2]{%
	\sbox\algocf@inoutbox{\KwSty{#2}\algocf@typo:}%
	\expandafter\ifx\csname InOutSizeDefined\endcsname\relax
	\newcommand\InOutSizeDefined{}%
	\sbox\algocf@inoutbox{\KwSty{#2}\algocf@typo\textbf{:}~}\setlength{\inoutindent}{\wd\algocf@inoutbox}%
	\else
	\ifdim\wd\algocf@inoutbox>\inoutsize%
	\sbox\algocf@inoutbox{\KwSty{#2}\algocf@typo\textbf{:}~}\setlength{\inoutindent}{\wd\algocf@inoutbox}%
	\fi%
	\fi
	\algocf@newcommand{#1}[1]{%
		\ifthenelse{\boolean{algocf@inoutnumbered}}{\relax}{\everypar={\relax}}%
		{\let\\\algocf@newinout\hangindent=\inoutindent\hangafter=1\KwSty{#2}\algocf@typo\textbf{:}~##1\par}%
		\algocf@linesnumbered
}}%

\makeatother
\begin{document}
	\title{Spectrum Sharing Among Multiple-Seller and Multiple-Buyer Operators of A Mobile Network: A Stochastic Geometry Approach}
	\author{Elaheh~Ataeebojd, Mehdi~Rasti,~\IEEEmembership{Senior Member,~IEEE}, Hossein~Pedram, and Pedro H. J. Nardelli,~\IEEEmembership{Senior~Member,~IEEE}%
		\thanks{This paper is published in part to the IEEE Wireless Communications and Networking Conference (WCNC), 2022 \cite{US2022}.}
		\thanks{E. Ataeebojd, M. Rasti, and H. Pedram are with the Department of Computer Engineering, Amirkabir University of Technology, Tehran, Iran, (e-mail: \{e.ataee, rasti, pedram\}@aut.ac.ir).}
		\thanks{Pedro H. J. Nardelli is with Lappeenranta-Lahti University of Technology, Lappeenranta, Finland, (e-mail: Pedro.Nardelli@lut.fi).}
		\thanks{This paper is partly supported by Academy of Finland via (a) FIREMAN consortium n.326270 as part of CHIST-ERA grant CHIST-ERA-17-BDSI-003, and (b) EnergyNet Research Fellowship n.321265/n.328869, and by Jane and Aatos Erkko Foundation via STREAM project. We would like to thank Dr. Hanna Niemel{\"a} for proofreading this paper.}
	}

	\IEEEtitleabstractindextext{%
		\begin{abstract}
			Sharing the spectrum among mobile network operators (MNOs) is a promising approach to improve the spectrum utilization and to increase the monetary income of MNOs. In this paper, we model a nonorthogonal spectrum sharing system for a large-scale cellular network where multiple seller MNOs lease their licensed sub-bands to several buyer MNOs. We first analyze the per-user expected rate and the per-MNO expected profit using stochastic geometry. Then, we formulate the joint problem of power control and licensed sub-band sharing to maximize the expected profit of all MNOs as a multiobjective optimization problem (MOOP) under the users' quality of service requirement and the nonnegative return on investment constraints. To transform the MOOP into a single objective form, we use a combination of the $\epsilon$-constraint and weighted sum methods. However, the transformed problem is nonconvex because of the presence of binary variables and nonconvex rate functions in the objective function and constraints. We address this problem by using a penalty function and approximating the nonconvex rate functions by a  constrained stochastic successive convex approximation method. Finally, the numerical results show the correctness and performance of the proposed algorithm under various conditions.
		\end{abstract}
		
		} 

\maketitle	
\IEEEdisplaynontitleabstractindextext
	
	\IEEEpeerreviewmaketitle
	
	\section{Introduction}\label{sec:introduction}
	\IEEEPARstart{T}{he} rising demand for various mobile services with different quality of services (QoS) results in an increasing need for spectrum resources in 5G and beyond \cite{Cisco}. However, the scarcity of the frequency spectrum and access to particular sub-bands respecting the license rights of different mobile network operators (MNOs) are the main challenges to meet these demands. Accordingly, pervasive sharing of the frequency spectrum is an efficient approach that leads to enhanced utilization of spectrum resources and meets the growing needs of spectrum resources. Sharing of the frequency spectrum has been considered in several schemes, including dynamic spectrum access \cite{Zhao2007} and cognitive radio \cite{Niyato2009}, \cite{Lu2020}. Nonetheless, most of these techniques are designed for sharing at a secondary level. In other words, a primary user with an exclusive right to a sub-band allows secondary users to use the same sub-band with a condition (i.e., when the primary user is idle or the received signal at the secondary users is under an accepted interference level of the primary user). Although such sharing enhances spectrum utilization, utilizing the spectrum depends on the decisions made between the primary and secondary users. 
	
	In recent years, there have been attempts to implement the spectrum sharing at the primary level by using a cognitive approach, with the focus on spectrum sharing among MNOs, e.g., co-primary \cite{Teng2014}, interoperator \cite{Cho2017}, and multioperator \cite{Luoto2017} sharing. Such spectrum sharing helps MNOs access more available frequency bands at a relatively small price and enhance spectrum utilization \cite{Cisco} --\cite{Matinmikko2014}. Further, it can be used to significantly improve the end users' QoS. Worldwide, MNOs, such as AT\&T, T-Mobile, and Verizon, have cooperated in spectrum sharing at the primary level \cite{Hou2021}.	However, spectrum sharing would cause additional costs to the MNOs in terms of direct utilization costs and potential performance degradation from an increase in interference levels \cite{GSMA}. Thus, the spectrum sharing scheme is efficient for all MNOs if the licensed bands are well harmonized to support their users' desired QoS, and the costs are kept reasonable \cite{Umar2017}. To achieve this, all MNOs must be careful to make correct choices of borrowing/leasing licensed bands. Because of the importance of guaranteeing QoS and effective utilization of the licensed spectrum, spectrum sharing is a significant trend in 5G and beyond; this is the focus of this paper.
	
	Different approaches for the improvement of spectrum utilization have been presented in the literature. For instance, the spectrum sharing approaches proposed in \cite{Cho2017}, \cite{Luoto2017}, \cite{Joshi2017} --\cite{Xiao2018}, \cite{Sanguanpuak2017} --\cite{Bairagi2019} are based on sharing the same spectrum among multiple MNOs. In contrast, in \cite{Asaduzzaman2018} --\cite{Gupta2016}, MNOs with a shortage of allocated spectrum can lease spectrum from other MNOs that have underutilized spectrum. For convenience, MNOs with a lack of allocated spectrum (underutilization spectrum) and their users are called buyer (seller) MNOs and buyer (seller) users, respectively. Furthermore, frequency bands may be shared either orthogonally or nonorthogonally among MNOs \cite{Irnich2013}. In orthogonal sharing, only one MNO is allowed to use a frequency band at any time. However, in nonorthogonal sharing, multiple MNOs are permitted to transmit simultaneously on the same frequency band.
	
	In \cite{Cho2017}, \cite{Luoto2017}, \cite{Joshi2017} --\cite{Xiao2018}, \cite{Sanguanpuak2017} --\cite{Bairagi2019}, the same spectrum is shared among multiple MNOs. More specifically, the licensed frequency bands are shared orthogonally among MNOs in \cite{Luoto2017} and \cite{Joshi2017}, whereas they are shared nonorthogonally in \cite{Cho2017} and \cite{Sanguanpuak2017}. In \cite{Joshi2017}, the problem of maximizing the weighted sum-rate for two MNOs is formally stated as a Nash bargaining game considering a dynamic network. According to Lyapunov optimization, both centralized and distributed dynamic algorithms are proposed in \cite{Joshi2017} to address the problem. In \cite{Luoto2017}, a decentralized approach is developed based on Gibbs sampling to allocate a fraction of the shared spectrum to each base station of MNOs while the long-term fairness is guaranteed. In \cite{Cho2017} and \cite{Sanguanpuak2017}, the authors first apply a stochastic geometry approach to analyze the long-term expected rate considering a large-scale network in which all base stations (BSs) are distributed based on the Poisson Point Process (PPP). After that, in \cite{Sanguanpuak2017}, a many-to-one stable matching game framework is adopted to model the problem of maximizing the social welfare of the MNOs. In contrast, in \cite{Cho2017}, a noncooperative game for interoperator device-to-device communications is established to maximize a constraint-based utility problem. Owing to the scarcity of the licensed spectrum and cost reduction, MNOs are also allowed to access the unlicensed frequency spectrum in \cite{Teng2017}, \cite{Xiao2018}, \cite{Zhang2017} --\cite{Bairagi2019}. The problem of maximizing the total utility based on a game-theoretical framework is formally stated in \cite{Teng2017} and \cite{Zhang2017} --\cite{Bairagi2019}. In \cite{Teng2017}, the problem is modeled as a repeated game respecting time-varying traffic. In \cite{Zhang2017}, the authors employ a multi-leader multi-follower Stackelberg game in which MNOs first set prices on each sub-band, and then, each user decides on them by applying a matching algorithm. In \cite{Bairagi2018}, a spectrum sharing scheme is proposed to maximize the quality of experience (measured in mean opinion score) of the users. A cooperative Nash bargaining game and a matching game are proposed to solve the problem under the users' QoS constraint in \cite{Bairagi2019}. In \cite{Xiao2018}, a network slicing game based on the overlapping coalition formation game is used to maximize the social welfare of MNOs so that the licensed and unlicensed spectrum resources are jointly distributed among MNOs according to the service demands and requirements of their users.
	
	Moreover, the spectrum sharing approach among seller and buyer MNOs is investigated as a solution in \cite{Asaduzzaman2018} --\cite{Gupta2016}. According to the spectrum price and the blocking probability, the authors of \cite{Asaduzzaman2018} formulate the problem of maximizing the profit of the buyer MNOs based on a  stochastic optimization framework. In \cite{Sanguanpuak2017Conf} --\cite{Gupta2016}, first, the signal-to-interference-plus-noise ratio (SINR) coverage probability and the expected rate are analyzed by employing a stochastic geometry approach. In \cite{Sanguanpuak2017Conf}, multiple seller MNOs lease their licensed sub-bands nonorthogonally to a single buyer MNO in indoor small cell networks. Further, a greedy algorithm is proposed to find the number of required licensed sub-bands for the buyer MNO under the QoS constraint in \cite{Sanguanpuak2017Conf}. The authors of \cite{Gupta2016} consider a system model consisting of one seller and one buyer MNO where the seller MNO shares its licensed band nonorthogonally with the buyer MNO regarding a limitation on the maximum interference that the seller MNO's users can tolerate. In addition,  a revenue model is presented for both MNOs based on this limitation.
	
	None of the aforementioned works consider the problem of maximizing the expected profit for all seller and buyer MNOs as a multiobjective optimization problem (MOOP). In this paper, the goal of the modeled multiobjective optimization problem (MOOP) is to derive a trade-off between maximizing the expected profit of the seller MNOs and maximizing the expected profit of the buyer MNOs. Furthermore, we employ a combination of the $\epsilon$-constraint and weighted sum methods to obtain a Pareto optimal front that satisfies the predefined conditions of Pareto optimality. Although the profit function is taken into account in \cite{Teng2017}, \cite{Xiao2018}, \cite{Zhang2017} --\cite{Bairagi2019}, these works consider a spectrum sharing scheme in which MNOs are in collaboration. In \cite{Asaduzzaman2018}, the problem of maximizing the profit function only for buyer MNOs is considered. In \cite{Cho2017} and \cite{Sanguanpuak2017}, the SINR coverage probability and the expected rate are analyzed by applying stochastic geometry in a co-primary system, while these performance metrics are investigated in a system with a single buyer MNO in \cite{Sanguanpuak2017Conf} --\cite{Gupta2016}. In \cite{Sanguanpuak2017Conf} --\cite{Gupta2016}, the interoperator interference is limited to seller MNOs and one buyer MNO. Thus, the impact of sharing spectrum for a BS of a buyer MNO on the neighboring ones of other buyer MNOs is not studied. In \cite{Cho2017}, \cite{Luoto2017}, \cite{Xiao2018} --\cite{Sanguanpuak2017Conf}, there is no power control over the downlink transmit power of the buyer MNOs. Consequently, the interference imposed on the seller users is unpredictable, resulting in a reduction in the performance of the seller networks.
	
	In this paper, by employing a stochastic geometry approach, we analyze the expected rate of users considering a large-scale cellular network with multiple seller and multiple buyer MNOs in which each MNO owns its BSs and users. Additionally, we assume that the downlink transmit power of the buyer MNOs, channel gains, and the number and location of users and BSs are random parameters. Moreover, inspired by \cite{Yan2018}, we limit the interference imposed on seller MNOs' users and adopt a distribution function for the downlink transmit power of buyer MNOs so that the interference imposed on seller MNOs' users does not violate a tolerable interference threshold. After that, we formally state the joint problem of power control and licensed sub-band sharing as a multiobjective optimization problem (MOOP) to maximize the expected profit of all MNOs so that the QoS requirement of the MNOs' users and the nonnegative return on investment of buyer MNOs are met. Although the stochastic analysis is more realistic, finding closed-form expressions for the objective function and constraints may be unfeasible. Therefore, designing an algorithm to solve the MOOP is challenging. To the best of the authors' knowledge, no previous studies have addressed the problem of maximizing the expected profit for both seller and buyer MNOs as a MOOP by a joint power control and licensed sub-band sharing in a stochastic environment by considering a maximum interference threshold on seller MNOs' users.
	
	The main contributions of this paper are summarized as follows:
	\begin{itemize}
		\item {In this paper, we extend the contributions of \cite{Sanguanpuak2017Conf} by proposing a more general framework containing several seller and buyer MNOs to study the impact of sharing sub-bands between seller and buyer MNOs. By employing stochastic geometry, we analytically obtain the expected rate of users and the expected profit of MNOs as performance metrics under random parameters, such as channel gains, the location of UEs and BSs, and the transmit power of buyer BSs. To achieve this, we introduce a power control strategy to justify the transmit power of buyer MNOs' BSs on the licensed sub-bands of the seller MNOs to ensure that the interference imposed on all users of each seller MNO is below the maximum interference threshold.
		}
		\item {By computing the expected rate of users and the expected profit of MNOs, we obtain an optimal solution for the stated joint power control and sub-band sharing problem. It is worth noting that we consider the joint problem of power control and sub-band sharing as a MOOP where we aim to maximize the expected profit of all MNOs under the QoS requirement of the MNOs' users and the nonnegative return on investment of the buyer MNOs.}
		\item {To address the stated MOOP, it is transformed into a single objective optimization problem by using a combination of the $\epsilon$-constraint and weighted sum methods. However, the binary sub-band sharing variables produce a disjoint feasible solution set that is an obstacle to solve the problem. To tackle this difficulty, we convert the binary variables into continuous ones by adding a penalty function to the objective function. Because of the presence of nonconvex expected rate functions in the objective function and the constraints, the transformed problem is still nonconvex. To address this challenge, we propose an algorithm based on the constrained stochastic successive convex approximation method by constructing a sequence of convex optimization problems that can be solved by using off-the-shelf software packages, such as CVX.}
		\item {We theoretically prove the convergence of the proposed algorithm to a stationary point, demonstrated by extensive numerical results. Additionally, numerical results show the performance of the proposed algorithm versus different varying parameters, such as the maximum interference threshold, the maximum transmit power of seller MNOs' BSs, the minimum data rate requirement of users, the number of sub-bands for seller MNOs, and the number of buyer MNOs.}
	\end{itemize}

	The remainder of this paper is organized as follows: Section \ref{Sys} introduces the system model and our assumptions. By employing stochastic geometry, we obtain the performance metrics in Section \ref{Ana}. Specifically, we compute the expected rate of the seller and buyer users and the expected profit for the seller and buyer MNOs. In Section \ref{Opt}, we formally state the problem of maximizing the expected profit of all MNOs as a MOOP. Then, we describe the proposed algorithm for solving the problem stated in Section \ref{Alg}. The numerical results and conclusions are presented in Section \ref{Res} and Section \ref{Con}, respectively.
	\section{System Model and Assumptions}\label{Sys}
	\subsection{System Model}
	We consider a downlink cellular wireless network in which a set of seller MNOs $\mathcal{S}$ $\left(|\mathcal{S}|=S\right)$ lease their frequency sub-bands to a set of buyer MNOs $\mathcal{B}$ $\left(|\mathcal{B}|=B\right)$. The set of all MNOs is denoted by $\mathcal{O} = \mathcal{S}\cup\mathcal{B}$. Each MNO owns an independent network consisting of its own BSs and users so that $\mathcal{F}_{k}$ and $\mathcal{U}_{k}$ represent the set of BSs and users for the MNO $k\in\mathcal{O}$, respectively. Moreover, it is assumed that each user is associated with the nearest BS of its MNO. The locations of BSs and users owned by the MNO $k\in \mathcal{O}$ are modeled by two independent PPPs, $\Phi_{k}$ with the intensity $\lambda_{k}$, and $\Psi_{k}$ with the intensity $\mu_{k}$.
	
	In this paper, we assume that each seller MNO $s\in\mathcal{S}$ owns a license for an orthogonal spectrum, which is divided into $L_{s}$ licensed sub-bands, the set of which is represented by $\mathcal{L}_{s}$ $\left(|\mathcal{L}_{s}| = L_{s}\right)$. In the set of licensed sub-bands $\mathcal{L}=\bigcup_{s\in\mathcal{S}} \mathcal{L}_{s}$ $\left(|\mathcal{L}| = L\right)$ for all seller MNOs, the sub-bands are also orthogonal to each other; in other words, there is no interference among the licensed sub-bands of different seller MNOs. Each seller MNO can simultaneously lease each of its licensed sub-bands to multiple buyer MNOs, and each buyer MNO may borrow multiple sub-bands from the seller MNOs. On the other hand, each seller MNO may reuse its licensed sub-bands to serve its users. Likewise, each buyer MNO may reuse its leased licensed sub-bands to serve its users. Therefore, in each licensed sub-band $l_{s}\in\mathcal{L}_{s}$ of seller MNO $s$, the corresponding seller users and several users of different buyer MNOs may get services simultaneously. Accordingly, in each sub-band $l_{s}\in\mathcal{L}_{s}$, a user of buyer MNO $b\in\mathcal{B}$ will experience interference simultaneously from three sets of BSs: from the BSs of its buyer MNO, the BSs of seller MNO $s$, and the BSs of the other buyer MNOs sharing the same sub-band. Similarly, a user of seller MNO $s$ will experience interference simultaneously from two sets of BSs: from the BSs of its seller MNO and the BSs of buyer MNOs sharing sub-band $l_{s}\in\mathcal{L}_{s}$.
	\begin{figure}[!t]	
		\centering
		\includegraphics[width=8.5 cm,height=6 cm]{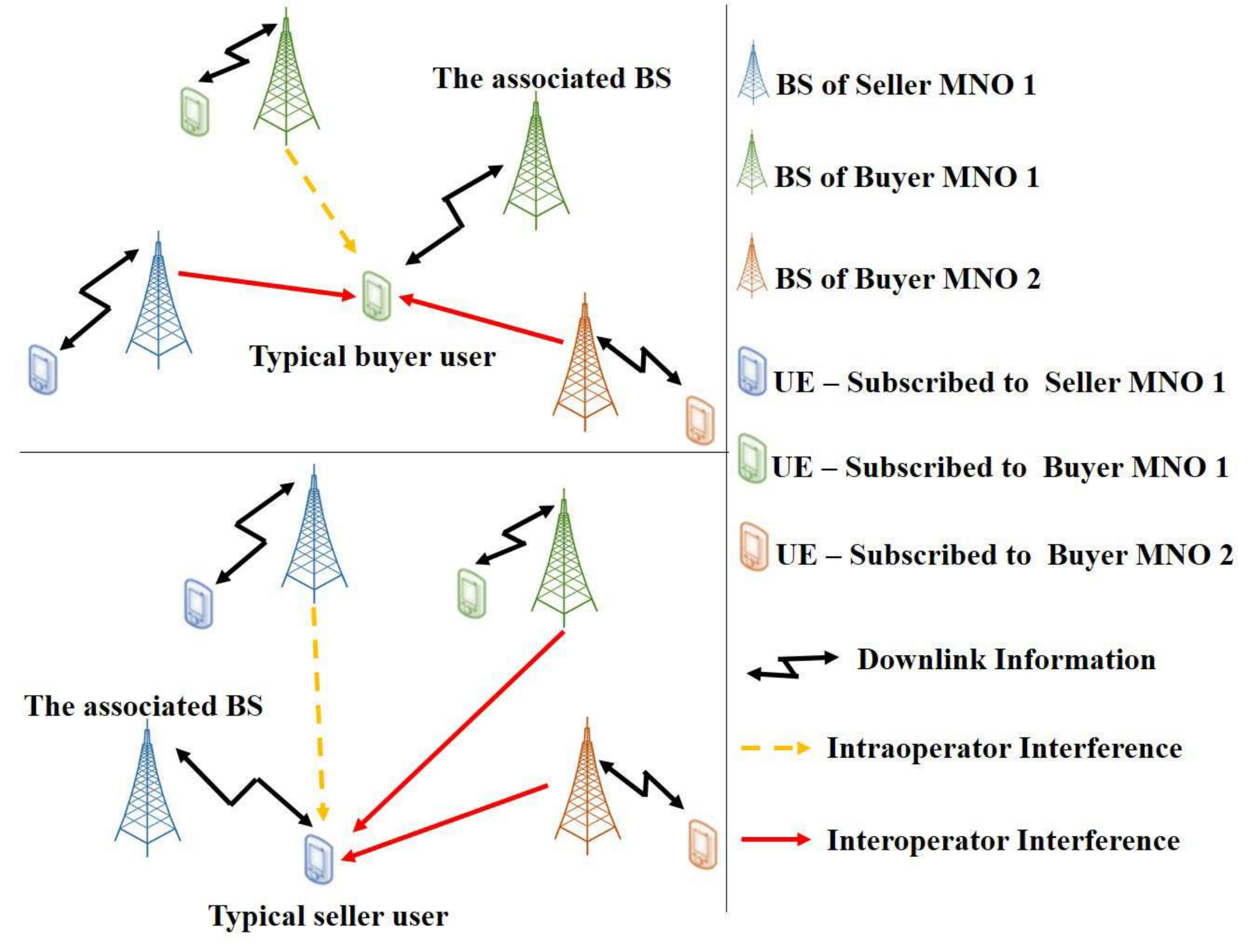}
		\caption{System model illustrating the SINR model for the typical buyer and seller user, where one seller MNO shares one of its licensed sub-bands with two buyer MNOs simultaneously. \label{sys}}
	\end{figure}
	\vspace{-1em}
	\subsection{Channel and SINR Model}
	Let us consider a downlink transmission between a typical user $\rm{UE}_{0}$ and its associated BS $\left(\text{BS}_{0}\right)$ for the MNO network $k\in\mathcal{O}$.\footnote{In the rest of this paper, $\rm{UE_{0}}$ and $\rm{BS_{0}}$ denote the typical user and its associated BS with which the typical user is associated, respectively.} We assume that the typical user $\rm{UE}_{0}$ is located at the origin, and the distance of $\text{BS}_{0}$ from $\rm{UE}_{0}$ is denoted by $x_{\rm{0}}$. Furthermore, the random variable $h_{0,l_{s}}$ is the channel fading gain between $\rm{UE}_{0}$ and $\rm{BS_{0}}$ on the sub-band $l_{s}$ of seller MNO $s$. In this work, we assume that all channel gains are modeled by the Rayleigh distribution model, which means that all fading variables are exponentially distributed with a mean 1 (i.e., $h_{0,l_{s}}\sim\exp\left(1\right)$). Moreover, $\alpha\:\left(>2\right)$ denotes the path-loss exponent.
	
	To represent sharing the sub-band $l_{s}$ between seller MNO $s$ and buyer MNO $b$, we define the binary variable $a_{l_{s},b}$; if the sub-band $l_{s}$ from seller MNO $s$ is leased to buyer MNO $b$, $a_{l_{s},b}=1$, otherwise, $a_{l_{s},b}=0$, where $\boldsymbol{\rm{A}}=\left[a_{l_{s},b}\right]_{L\times B}^{T}$ denotes sharing sub-bands among all seller and buyer MNOs. In addition, the vector $\boldsymbol{\rm{P}}=\left[p_{s,l_{s}}\right]_{L}^{T}$ represents the downlink transmit power of the seller MNOs on licensed sub-bands, in which the element $p_{s,l_{s}}$ denotes the transmit power of each BS of seller MNO $s$ on the sub-band $l_{s}$.
	
	The downlink received SINR at a typical user $\rm{UE}_{0}$ of seller MNO $s$ on the sub-band $l_{s}$ (see Fig. \ref{sys}) is determined by
	\begin{equation}
	\begin{aligned}
	\gamma_{s,l_{s}}\left(\boldsymbol{\rm{A}},\boldsymbol{\rm{P}}\right)=\frac{p_{s,l_{s}}h_{0,l_{s}}{x}_{0}^{-\alpha}}{I_{s,l_{s}}\left(\boldsymbol{\rm{A}},\boldsymbol{\rm{P}}\right)+\sigma^{2}},
	\end{aligned}
	\end{equation}
	where ${\sigma}^2$ is the noise power, and $I_{s,l_{s}}\left(\boldsymbol{\rm{A}},\boldsymbol{\rm{P}}\right)$ is the interference experienced by this typical user from all BSs that share the sub-band $l_{s}$, which is calculated by $I_{s,l_{s}}\left(\boldsymbol{\rm{A}},\boldsymbol{\rm{P}}\right)=\sum_{f \in \mathcal{F}_{s}\setminus\{0\}} p_{s,l_{s}} h_{f,l_{s}} x_{f}^{-\alpha} + \sum_{b \in \mathcal{B}} \sum_{f\in\mathcal{F}_{b}} a_{l_{s},b} p_{b,l_{s}} h_{f,l_{s}}\allowbreak x_{f}^{-\alpha}$, where $x_{f}$ is the distance of this typical user from BS $f$, and $h_{f,l_{s}}$ is the channel fading gain between this typical user and BS $f$ on the sub-band $l_{s}$ of seller MNO $s$. As seen from $I_{s,l_{s}}\left(\boldsymbol{\rm{A}},\boldsymbol{\rm{P}}\right)$, the first term describes the intraoperator interference, while the second term represents the interoperator interference from buyer MNOs.
	
	Likewise, the downlink received SINR at a typical user $\rm{UE}_{0}$ of buyer MNO $b$ on the shared sub-band $l_{s}$ (see Fig. \ref{sys}) is expressed as
	\begin{equation}\label{SINR_m}
	\begin{aligned} 
	\gamma_{b,l_{s}}\left(\boldsymbol{\rm{A}},\boldsymbol{\rm{P}}\right)=\dfrac{p_{b,l_{s}}h_{0,l_{s}}{x}_{0}^{-\alpha}}{I_{b,l_{s}}\left(\boldsymbol{\rm{A}},\boldsymbol{\rm{P}}\right)+\sigma^{2}},
	\end{aligned}
	\end{equation}
	where $p_{b,l_{s}}$ is the transmit power of each BS of buyer MNO $b$ on the sub-band $l_{s}$ of seller MNO $s$, and $I_{b,l_{s}}\left(\boldsymbol{\rm{A}},\boldsymbol{\rm{P}}\right)$ is the interference experienced by this typical user from all BSs that share the sub-band $l_{s}$, which is obtained by $I_{b,l_{s}}\left(\boldsymbol{\rm{A}},\boldsymbol{\rm{P}}\right) = \sum_{f\in\mathcal{F}_{b}\setminus\{0\}} a_{l_{s},b} p_{b,l_{s}} h_{f,l_{s}} x_{f}^{-\alpha} + \sum_{f \in \mathcal{F}_{s}}\allowbreak p_{s,l_{s}} h_{f,l_{s}} x_{f}^{-\alpha} + \sum_{b^{'} \in \mathcal{B}\setminus\{b\}} \sum_{f\in\mathcal{F}_{b^{'}}} a_{l_{s},b^{'}} p_{b^{'},l_{s}} h_{f,l_{s}}\allowbreak x_{f}^{-\alpha}$. As seen from $I_{b,l_{s}}\left(\boldsymbol{\rm{A}},\boldsymbol{\rm{P}}\right)$, the first term relates to the intraoperator interference, while the second and third terms express the interoperator interference from seller MNO $s$ and other buyer MNOs, respectively.
	
	Based on Shannon's formula, the instantaneous rates of the considered typical users of seller MNO $s$ and buyer MNO $b$ on the sub-band $l_{s}$ is given by $r_{s,l_{s}}\left(\boldsymbol{\rm{A}},\boldsymbol{\rm{P}}\right)=\log_{2}\left(1+\gamma_{s,l_{s}}\left(\boldsymbol{\rm{A}},\boldsymbol{\rm{P}}\right)\right)$ and $r_{b,{l_{s}}}\left(\boldsymbol{\rm{A}},\boldsymbol{\rm{P}}\right)=a_{l_{s},b}\log_{2}\left(1+{\gamma_{b,{l_{s}}}}\left(\boldsymbol{\rm{A}},\boldsymbol{\rm{P}}\right)\right)$, respectively. Accordingly, the total instantaneous rates of the typical users of seller MNO $s$ and buyer MNO $b$ are obtained by $r_{s}\left(\boldsymbol{\rm{A}},\boldsymbol{\rm{P}}\right)=\sum_{l_{s}\in\mathcal{L}_{s}}^{}r_{s,l_{s}}\left(\boldsymbol{\rm{A}},\boldsymbol{\rm{P}}\right)$ and $r_{b}\left(\boldsymbol{\rm{A}},\boldsymbol{\rm{P}}\right)=\sum_{s\in\mathcal{S}}^{}\sum_{l_{s}\in\mathcal{L}_{s}}^{}r_{b,{l_{s}}}(\boldsymbol{\rm{A}},\boldsymbol{\rm{P}})$, respectively.
	\subsection{Power Control Strategy} \label{Power_Strategy}
	Inspired by the approach taken in \cite{Yan2018}, each BS of buyer MNO $b$ is limited to transmit at a certain power $p_{b,l_{s}}$ on the sub-band $l_{s}\in\mathcal{L}_{s}$ to ensure that the total interference imposed on each user of seller MNO $s$ is below the maximum interference threshold $\zeta_{s}$. Thus, given the maximum interference threshold $\zeta_{s}$, for any user $i\in\mathcal{U}_{s}$ located at ${\rm{y}}_{i}$, we have $\sum_{b\in\mathcal{B}}\sum_{f\in\mathcal{F}_{b}} p_{b,l_{s}}h_{i,f,l_{s}}{||{\rm{y}}_{i}-{\rm{x}}_{f}||}^{-\alpha}\leq \zeta_{s}$, where $h_{i,f,l_{s}}$ is the channel gain between the user $i$ and BS $f$ on the sub-band $l_{s}$, and $\text{x}_{f}$ is the location of BS $f$. Due to the fact that the total interference distribution is not known in closed form  when buyer BSs are distributed based on the PPP \cite{Haenggi2009}, we approximate the total interference experienced by each seller user with its strongest interferer \cite{Li2016}. Thus, we have $\sum_{b\in\mathcal{B}}\sum_{f\in\mathcal{F}_{b}} p_{b,l_{s}}h_{i,f,l_{s}}{||{\rm{y}}_{i}-{\rm{x}}_{f}||}^{-\alpha}\leq \zeta_{s} \approx \max_{\substack{b\in\mathcal{B}\\f\in\mathcal{F}_{b}}} p_{b,l_{s}}h_{i,f,l_{s}}{||{\rm{y}}_{i}-{\rm{x}}_{f}||}^{-\alpha}\leq \zeta_{s}$. Hence, for any user $i\in\mathcal{U}_{s}$ located at ${\rm{y}}_{i}$, we have $ p_{b,l_{s}}h_{i,f,l_{s}}{||{\rm{y}}_{i}-{\rm{x}}_{f}||}^{-\alpha}\leq \zeta_{s}$, for all $f\in\mathcal{F}_{b}$ and for all $b\in\mathcal{B}$. Accordingly, it is sufficient to satisfy the interference constraint for such a user of seller MNO $s$ that experiences the maximum interference power from each BS of buyer MNO $b$. Thus, given an arbitrary BS $f$ of buyer MNO $b$, we have $\max_{{\rm{y}}_{i} \in \Psi_{s}} {p_{b,l_{s}}h_{i,f,l_{s}}{||{\rm{y}}_{i}-{\rm{x}}_{f}||}^{-\alpha}}\leq\zeta_{s},$ for all $f\in\mathcal{F}_{b}$ and for all $b\in\mathcal{B}$. By doing this, the interference constraint is surely satisfied for all other users of seller MNO $s$. Let us denote the largest interference channel gain on the sub-band $l_{s}$ associated with each BS of buyer MNO $b$ as ${H}_{b,l_{s}}$. It is defined as ${H}_{b,l_{s}} = \max_{{\rm{y}}_{i} \in \Psi_{s}} {h_{i,f,l_{s}}{||{\rm{y}}_{i}-{\rm{x}}_{f}||}^{-\alpha}}$, for all $f\in\mathcal{F}_{b}$ and for all $b\in\mathcal{B}$. If each BS of buyer MNO $b$ transmits with the maximum allowable power on the sub-band $l_{s}\in\mathcal{L}_{s}$, $p_{b,l_{s}}$ is given by
	\begin{equation}
	\begin{aligned} \label{Power}
	p_{b,l_{s}}=\zeta_{s}/H_{b,l_{s}}.
	\end{aligned}
	\end{equation}
	\section{Analysis of Performance Metrics}\label{Ana}
	In this paper, we aim to maximize the expected profit of all seller and buyer MNOs simultaneously. To achieve this, we formally state this problem as a MOOP subject to the users' QoS requirement and the constraints of nonnegative return on investment. In order to formulate this problem, we need to calculate the expected profit of each buyer and seller MNO, which depends on the expected rate of their users. Consequently, we need to obtain the distribution of $H_{b,l_{s}}$ and $p_{b,l_{s}}$ to derive the expected rate of users. In what follows, we first derive the distribution of $H_{b,l_{s}}$ and $p_{b,l_{s}}$. After that, we derive the expected rate of each seller and buyer user under randomness of channel fading, distance geometry, and transmit power of buyer MNOs' BSs. Finally, based on the expected rate of users, the expected profit of MNOs is obtained. The expected rate of users and the expected profit of MNOs are regarded as the performance metrics.
	
	\textit{1) Distribution of $H_{b,l_{s}}$ and $p_{b,l_{s}}$:} Because channel gains and the distance of users from BSs are random variables, $H_{b,l_{s}}$ and $p_{b,l_{s}}$ are also random variables. In the following lemmas, the distributions of $H_{b,l_{s}}$ and $p_{b,l_{s}}$ are derived \cite{Yan2018}.
	
	\begin{lem} \label{lemma_1}
		The cumulative distribution function (CDF) and the probability density function (PDF) of $H_{b,l_{s}}$ are expressed respectively as 
		\begin{equation}
		\begin{aligned}
		&F_{H_{b,l_{s}}}\left(z\right)=\exp\left(\frac{-\pi\mu_{s}}{z^{2/\alpha}}\Gamma\left(1+\frac{2}{\alpha}\right)\right)\label{CDF_H},
		\end{aligned}
		\end{equation}	
		and
		\begin{equation}
		\begin{aligned}
		&f_{H_{b,l_{s}}}\left(z\right)= \frac{2\pi\mu_{s}\Gamma\left(1+\frac{2}{\alpha}\right)}{\alpha z^{1+\frac{2}{\alpha}}} \exp\left(\frac{-\pi\mu_{s}}{z^{2/\alpha}}\Gamma\left(1+\frac{2}{\alpha}\right)\right), \label{PDF_H}
		\end{aligned}
		\end{equation}
		where $\Gamma(\kappa)$ is the complete Gamma function calculated as $\Gamma(\kappa)=\int_{0}^{+\infty}{t^{\kappa-1}e^{-t}\;{\rm{d}}t}$.
	\end{lem}
	\begin{lem} \label{lemma_2}
		The CDF and PDF of $p_{b,l_{s}}$ are expressed respectively as
		\begin{equation}
		\begin{aligned}
		&F_{p_{b,l_{s}}}\left(z\right)=1-\exp\left(\frac{-\pi\mu_{s}z^{2/\alpha}}{\zeta_{s}^{2/\alpha}}\Gamma\left(1+\frac{2}{\alpha}\right)\right)\label{CDF_P},
		\end{aligned}
		\end{equation}
		and
		\begin{equation}\small
		\begin{aligned}
		f_{p_{b,l_{s}}}(z)=& \frac{2\pi\mu_{s}z^{2/\alpha-1}\Gamma\left(1+\frac{2}{\alpha}\right)}{\alpha \zeta_{s}^{\frac{2}{\alpha}}}\exp\left(\frac{-\pi\mu_{s}z^{2/\alpha}}{\zeta_{s}^{2/\alpha}}\Gamma\left(1+\frac{2}{\alpha}\right)\right). \label{PDF_P}
		\end{aligned}
		\end{equation}
	\end{lem}
	\textit{2) Expected rate of users:} Without loss of generality, we analyze the expected rate for a typical user $\rm{UE}_{0}$ of a seller MNO and a buyer MNO considering a network in which multiple seller MNOs lease their sub-bands to multiple buyer MNOs. Similarly, the expected rate for other buyer and seller MNOs' users is mathematically analyzed.
	
	Similar to the steps taken in \cite{Andrews2011}, we calculate the expectation of $r_{b}\left(\boldsymbol{\rm{A}},\boldsymbol{\rm{P}}\right)$ and $r_{s}(\boldsymbol{\rm{A}},\boldsymbol{\rm{P}})$ with respect to the channel gains, transmit powers, and distance geometry:
	\begin{equation}
	\begin{aligned} \label{r_m_}
	\overline{r}_{b}\left(\boldsymbol{\rm{A}},\boldsymbol{\rm{P}}\right) = \mathbb{E}\big[r_{b}\left(\boldsymbol{\rm{A}},\boldsymbol{\rm{P}}\right)\big],
	\end{aligned}
	\end{equation}
	and
	\begin{equation}
	\begin{aligned} \label{r_n_}
	\overline{r}_{s}\left(\boldsymbol{\rm{A}},\boldsymbol{\rm{P}}\right) = \mathbb{E}\big[r_{s}\left(\boldsymbol{\rm{A}},\boldsymbol{\rm{P}}\right)\big],
	\end{aligned}
	\end{equation}
	respectively. Specifically, the expected rate for a typical user of seller MNO $s$ and buyer MNO $b$ is analytically presented in the following lemmas.\footnote{In Section \ref{Res}, the analysis results provided in Lemmas \ref{thm_1} and \ref{thm_2} are used to obtain the optimal solution.}
	\begin{lem} \label{thm_1}
		The expected rate for a typical user of buyer MNO $b$ is
		\begin{equation} \label{r_m}
		\begin{aligned}
		&{\overline{r}}_{b}(\boldsymbol{\rm{A}},\boldsymbol{\rm{P}})=\mathbb{E}\Big[\sum_{s\in\mathcal{S}}\sum_{l_{s}\in \mathcal{L}_{s}}a_{l_{s},b}\log\left(1+{\gamma_{b,{l_{s}}}}\right)\Big]=\\
		&\sum_{s\in\mathcal{S}}\sum_{l_{s}\in \mathcal{L}_{s}}\int_{0}^{\infty}\int_{0}^{+\infty}a_{l_{s},b}\:\pi\lambda_{b}K\left(\mu_{s}\right)\exp\{-\left(2^{t}-1\right)\sigma^{2}z^{\alpha/2}\\
		&-\pi(\lambda_{b}K(\mu_{s}){\left(2^{t}-1\right)}^{2/\alpha}\rho\left(\alpha,(2^{t}-1)\right)+\lambda_{s}p_{s,l_{s}}^{2/\alpha}{\left(2^{t}-1\right)}^{2/\alpha}\\
		& \rho\left(\alpha,\infty\right)+\sum_{b^{'} \in \mathcal{B}\setminus\{b\}}a_{l_{s},b^{'}}\lambda_{b^{'}}K(\mu_{s}){\left(2^{t}-1\right)}^{2/\alpha}\rho\left(\alpha,\infty\right) + \lambda_{b}\\
		&K(\mu_{s}))z\}\:{\rm{d}}z\:{\rm{d}}t,
		\end{aligned}
		\end{equation}
		where $\rho\left(\alpha,T\right)=\int_{{T}^{-2/\alpha}}^{+\infty}\frac{1}{1+\nu^{\alpha/2}} {\rm{d}}\nu$, $\rho\left(\alpha,\infty\right)=\int_{0}^{+\infty}\frac{1}{1+\nu^{\alpha/2}}{\rm{d}}\nu$, and $K(\mu_{s})=\mathbb{E}_{p_{b,l_{s}}}\left[{p_{b,l_{s}}}^{2/\alpha}\right]$.
	\end{lem}
	\begin{proof}
		The expected rate for a typical user of buyer MNO $b$ is given by
		\begin{equation}
		\begin{aligned}
		\overline{r}_{b} &= \mathbb{E}\Big[\sum_{s\in\mathcal{S}}\sum_{l_{s}\in\mathcal{L}_{s}}a_{l_{s},b}\log(1+\gamma_{b,{l_{s}}})\Big] \\
		&=\sum_{s\in\mathcal{S}}\sum_{l_{s}\in\mathcal{L}_{s}}a_{l_{s},b} \mathbb{E}\Big[\log(1+\gamma_{b,{l_{s}}})\Big].
		\end{aligned}
		\end{equation}
		
		To obtain $\overline{r}_{b}$, it is important to accurately calculate $\overline{r}_{b,l_{s}} =\mathbb{E}[\log(1+\gamma_{b,{l_{s}}})]$. From the fact that the expectation of any positive random variable $x$ can be given by $\mathbb{E}[x]=\int_{0}^{\infty}\mathbb{P}(x>t)\;{\rm{d}}t$, $\overline{r}_{b,l_{s}}$ can be written as
		\vspace{-0.5 em}
		\begin{equation}
		\begin{aligned} \label{rate_1}
		&\overline{r}_{b,l_{s}} =\mathbb{E}\Big[\int_{0}^{\infty}\mathbb{P}\left(\log(1+\gamma_{b,{l_{s}}})>t\right)\;{\rm{d}}t\Big]\\
		&=\mathbb{E}\Big[\int_{0}^{\infty}\mathbb{P}\left(\gamma_{b,{l_{s}}}>2^{t}-1\right)\;{\rm{d}}t\Big]\\
		&=\mathbb{E}\Big[\int_{0}^{\infty} \mathbb{P}\left(\frac{p_{b,l_{s}}h_{0,l_{s}}{x_{0}^{-\alpha}}}{I_{b,l_{s}}+\sigma^{2}}>2^{t}-1\right)\;{\rm{d}}t\Big]\\
		&\overset{(a)}{=}\mathbb{E}\Big[\int_{0}^{\infty} \int_{0}^{\infty}\mathbb{P}\left(\frac{p_{b,l_{s}}h_{0,l_{s}}{x_{0}^{-\alpha}}}{I_{b,l_{s}}+\sigma^{2}}>2^{t}-1\right)f_{b}(x)\;{\rm{d}}x\;{\rm{d}}t\Big]\\
		&\overset{(b)}{=}\mathbb{E}\Big[\int_{0}^{\infty} \int_{0}^{\infty}{(p_{b,l_{s}})}^{2/\alpha}\mathbb{P}\left(\frac{h_{0,l_{s}}{u}^{-\alpha}}{I_{b,l_{s}}+\sigma^{2}}>2^{t}-1\right)f_{b}(u)\;{\rm{d}}u\;{\rm{d}}t\Big]\\
		&\overset{(c)}{=}\int_{0}^{\infty}\int_{0}^{\infty}K(\mu_{s})\mathbb{E}\Big[\exp\{-u^{\alpha}(2^{t}-1)(\sigma^{2}+I_{b,l_{s}})\}f_{b}(u)\;\\
		&{\rm{d}}u\;{\rm{d}}t\Big]\overset{(d)}{=}\int_{0}^{\infty}\int_{0}^{\infty}K(\mu_{s})\exp\left(-u^{\alpha}(2^{t}-1)\sigma^{2}\right)\\
		&\mathcal{L}_{I_{b,l_{s}}}\left(u^{\alpha}(2^{t}-1)\right) f_{b}(u)\;{\rm{d}}u\;{\rm{d}}t,
		\end{aligned}
		\end{equation}
		where $(a)$ is based on the fact that the typical user of buyer MNO $b$ is associated with the nearest BS of its MNO network, in which the PDF of the distance $x$ between the typical user and its nearest BS is computed by
		\begin{align} \label{F}
		f_{b}(x) = 2\pi\lambda_{b}x\exp\left(-\pi\lambda_{b}x^{2}\right),
		\end{align}
		$(b)$ is obtained by using the transformation $x = {(p_{b,l_{s}})}^{1/\alpha}u$; $(c)$ is due to (\ref{Power}) and the definition of $K(\mu_{s})$; and $(d)$ is from the moment generating function of $h_{0,l_{s}}$.
		
		$\mathcal{L}_{I_{b,l_{s}}}(u^{\alpha}(2^{t}-1))$ is the Laplace transform of $I_{b,l_{s}}$, given by
		\begin{equation}
		\begin{aligned} \label{Laplace}
		&\mathcal{L}_{I_{b,l_{s}}}\left(u^{\alpha}(2^{t}-1)\right) = 
		\exp\Big(-\pi\lambda_{b}K(\mu_{s})(2^t-1)^{2/\alpha}u^{2}\\
		&\rho(\alpha,(2^t -1))\Big)
		\exp\Big(-\pi\lambda_{s}p_{s,l_{s}}^{2/\alpha}(2^t-1)^{2/\alpha}u^{2}\rho(\alpha,\infty)\Big)\\
		&\prod_{b^{'}\in\mathcal{B}\setminus\{b\}}^{}\exp\Big(-\pi a_{l_{s},b}\lambda_{b^{'}}K(\mu_{s})(2^t-1)^{2/\alpha}u^{2}\rho(\alpha,\infty)\Big).
		\end{aligned}
		\end{equation}
		The computation of $\mathcal{L}_{I_{b,l_{s}}}(u^{\alpha}(2^{t}-1))$ is straightforward and omitted owing to page limitations.
		
		By substituting (\ref{F}) and (\ref{Laplace}) into (\ref{rate_1}) and using the transformation $u^{2} = z$, the proof of lemma is completed.
	\end{proof}
	\begin{lem} \label{thm_2}
		The expected rate for a typical user of seller MNO $s$ is
		\begin{equation}\label{r_n}
		\begin{aligned}
		&{\overline{r}}_{s}(\boldsymbol{\rm{A}},\boldsymbol{\rm{P}})=\mathbb{E}\Big[\sum_{l_{s}\in \mathcal{L}_{s}}\log\left(1+{\gamma}_{l_{s}}\right)\Big]=\\
		&\sum_{l_{s}\in \mathcal{L}_{s}}\int_{0}^{\infty}\int_{0}^{+\infty}\pi\lambda_{s}p_{s,l_{s}}^{2/\alpha}\exp\{-\left(2^{t}-1\right)\sigma^{2}z^{\alpha/2}-\pi (\lambda_{s}p_{s,l_{s}}^{2/\alpha}\\
		&\left(2^{t}-1\right)^{2/\alpha}\rho\left(\alpha,\left(2^{t}-1\right)\right)+\sum_{b\in\mathcal{B}}a_{l_{s},b}\lambda_{b}K(\mu_{s}){\left(2^{t}-1\right)}^{2/\alpha}\\
		&\rho\left(\alpha,\infty\right)+\lambda_{s}p_{s,l_{s}}^{2/\alpha})z\}\:{\rm{d}}z\:{\rm{d}}t.
		\end{aligned}
		\end{equation}
	\end{lem}
	\begin{proof}
		The expected rate for a typical user of seller MNO $s$ is expressed as
		\begin{equation}
		\begin{aligned}
		\overline{r}_{s} &= \mathbb{E}\Big[\sum_{l_{s}\in\mathcal{L}_{s}}\log(1+\gamma_{s,l_{s}})\Big] = \sum_{l_{s}\in\mathcal{L}_{s}} \mathbb{E}\Big[\log(1+\gamma_{s,l_{s}})\Big].
		\end{aligned}
		\end{equation}
		To obtain $\overline{r}_{s}$, it is important to accurately calculate $\overline{r}_{s,l_{s}} =\mathbb{E}[\log(1+\gamma_{s,l_{s}})]$. Similar to the computations for $\overline{r}_{b,l_{s}}$, $\overline{r}_{s,l_{s}}$ can be calculated. The only difference is that the computations for the buyer and seller MNOs are interchanged, and the downlink transmit power of the seller MNOs is deterministic.
	\end{proof}
	\textit{3) Expected profit of MNOs:} The expected profit of buyer MNO $b$, denoted by $u_b$, at the end of the duration interval time $D$ (investment time) is obtained by
	\begin{equation}
	\begin{aligned} \label{u_b}
	u_{b}(\boldsymbol{\rm{A}},\boldsymbol{\rm{P}}) = \nu_{b}(\boldsymbol{\rm{A}},\boldsymbol{\rm{P}}) - c_{b}(\boldsymbol{\rm{A}}),
	\end{aligned}
	\end{equation}
	where $\nu_{b}(\boldsymbol{\rm{A}},\boldsymbol{\rm{P}})$ and $c_{b}(\boldsymbol{\rm{A}})$ are the expected revenue and cost functions for buyer MNO $b$, respectively, which are obtained as explained in the following.
	
	$\nu_{b}(\boldsymbol{\rm{A}},\boldsymbol{\rm{P}})$ is a linear function of $\overline{r}_{b}(\boldsymbol{\rm{A}},\boldsymbol{\rm{P}})$ and expressed as
	\begin{equation}
	\begin{aligned} \label{rev_m}
	\nu_{b}(\boldsymbol{\rm{A}},\boldsymbol{\rm{P}}) = \delta D {N}_{b} {\overline{r}}_{b}(\boldsymbol{\rm{A}},\boldsymbol{\rm{P}}),
	\end{aligned}
	\end{equation}
	where $\delta$ represents a monthly price per $1 \rm{bps}/\rm{Hz}$, and ${N}_{b}$ is the expected number of users subscribing to MNO $b$ in a coverage area. It is calculated as ${N}_{b} = \pi r^{2}\mu_{b}$, in which $r$ is the radius of the coverage area. Given the monthly price $\delta$, each buyer user must pay a monthly cost to its MNO based on its expected rate obtained by (\ref{r_m_}). According to the number of sub-bands leased by buyer MNO $b$ and the price coefficient $\varphi_{s,b}$ (the price that buyer MNO $b$ must pay to seller MNO $s$ for each sub-band leased), the cost function $c_{b}(\boldsymbol{\rm{A}})$ is determined by
	\begin{equation}
	\begin{aligned} \label{Cost}
	c_{b}(\boldsymbol{\rm{A}})=\sum_{s\in\mathcal{S}}^{}\sum_{l_{s}\in\mathcal{L}_{s}}^{}\varphi_{s,b}\:a_{l_{s},b}.
	\end{aligned}
	\end{equation}
	
	Likewise, inspired by \cite{Gupta2016} and \cite{Cano2016}, the expected profit of seller MNO $s$, denoted by $u_{s}$, at the end of investment time $D$ is calculated by
	\begin{equation}
	\begin{aligned} \label{u_n}
	u_{s}(\boldsymbol{\rm{A}},\boldsymbol{\rm{P}}) =  \nu_{s}(\boldsymbol{\rm{A}},\boldsymbol{\rm{P}}) - c_{s},
	\end{aligned}
	\end{equation}
	where $\nu_{s}(\boldsymbol{\rm{A}},\boldsymbol{\rm{P}})$ and $c_{s}$ are the expected revenue and cost functions for seller MNO $s$, respectively, which are obtained as explained in the following.
	
	We assume that $\nu_{s}^{1}(\boldsymbol{\rm{A}},\boldsymbol{\rm{P}})$ is the revenue obtained by seller MNO $s$ from the payments of its own users, that is,
	\begin{equation}
	\begin{aligned}
	\nu_{s}^{1}(\boldsymbol{\rm{A}},\boldsymbol{\rm{P}})=\delta D N_{s} {\overline{r}}_{s}(\boldsymbol{\rm{A}},\boldsymbol{\rm{P}}),
	\end{aligned}
	\end{equation}
	and $\nu_{s}^{2}(\boldsymbol{\rm{A}})$ is the revenue obtained by seller MNO $s$ from leasing its sub-bands to the buyer MNOs,  given by
	\begin{equation}
	\begin{aligned}
	\nu_{s}^{2}(\boldsymbol{\rm{A}})=\sum_{l_{s}\in\mathcal{L}_{s}}^{}\sum_{b\in\mathcal{B}}^{}\varphi_{s,b}\:a_{l_{s},b}.
	\end{aligned}
	\end{equation}
	Consequently, $\nu_{s}(\boldsymbol{\rm{A}},\boldsymbol{\rm{P}})$ is defined as 
	\begin{equation}
	\begin{aligned} \label{v_n}
	\nu_{s}(\boldsymbol{\rm{A}},\boldsymbol{\rm{P}}) =  \nu_{s}^{1}(\boldsymbol{\rm{A}},\boldsymbol{\rm{P}}) + \nu_{s}^{2}(\boldsymbol{\rm{A}}).
	\end{aligned}
	\end{equation}
	
	Let $\phi_{s}$ denote the license price per sub-band paid by seller MNO $s$ to regulators. The cost function $c_{s}$ is determined by
	\begin{equation}
	\begin{aligned} \label{Cost_s}
	c_{s}=\phi_{s}\:L_{s}.
	\end{aligned}
	\end{equation}
	\vspace{-2.0 em}
	\section{Optimization Problem Formulation}\label{Opt}
	In this section, we formally state the joint problem of power control and sub-band sharing for maximizing the expected profit of all MNOs simultaneously subject to the users' QoS requirement and nonnegative return on investment constraints as follows:
	\begin{equation} \label{Problem_2}
	\begin{aligned}
	&\max\limits_{\boldsymbol{\rm{A}},\boldsymbol{\rm{P}}}
	&\quad
	& \{u_{k}(\boldsymbol{\rm{A}},\boldsymbol{\rm{P}})\}_{\forall k \in \mathcal{O}}\\
	& \text{s.t.} &
	& \text{C1}:{\overline{r}}_{k}(\boldsymbol{\rm{A}},\boldsymbol{\rm{P}})\geq {r}^{\rm{th}},
	&& \forall k \in \mathcal{O},\\
	&&&\text{C2}:u_{b}(\boldsymbol{\rm{A}},\boldsymbol{\rm{P}})\geq 0,
	&& \forall b \in \mathcal{B}, \\
	&&& \text{C3}:\sum_{l_{s}\in \mathcal{L}_{s}}^{}a_{l_{s},b} \leq N_{s}^{\rm{B}},
	&& \forall s \in \mathcal{S}, \forall b \in \mathcal{B},\\
	&&& \text{C4}:\sum_{b\in \mathcal{B}}^{}a_{l_{s},b} \leq N_{s}^{\rm{L}},
	&& \forall s \in \mathcal{S}, \forall l_{s} \in \mathcal{L}_{s},\\
	&&& \text{C5}:0 \leq p_{s,l_{s}}\leq P^{\rm{max}},
	&&\forall s\in \mathcal{S}, \forall l_{s} \in \mathcal{L}_{s},\\
	&&& \text{C6}:a_{l_{s},b}\in \{0,1\},
	&&\forall s\in \mathcal{S}, \forall b\in \mathcal{B}, \forall l_{s} \in \mathcal{L}_{s}.
	\end{aligned}
	\end{equation}
	In the MOOP (\ref{Problem_2}), constraint $\text{C1}$ guarantees the required minimum expected rate for each user of the MNOs, which is denoted by ${r}^{\rm{th}}$. Further, constraint $\text{C2}$ ensures that each buyer MNO obtains a nonnegative return on its investment. Moreover, constraint $\text{C3}$ determines that each buyer MNO $b$ can borrow  $N_{s}^{\rm{B}}$ sub-bands at most from the seller MNO $s$. Moreover, $\text{C4}$ denotes that each sub-band $l_{s}$ of seller MNO $s$ can be leased by $N_{s}^{\rm{L}}$ buyer MNOs at most. In addition, constraint $\text{C5}$ guarantees the maximum and minimum transmit power limitation on each sub-band of the seller MNOs, in which $P^{\text{max}}$ is the maximum allowable transmit power. Furthermore, the binary nature of the sub-band sharing variable is implied in $\text{C6}$.
	\vspace{-0.6 em}
	\section{Proposed Algorithm}\label{Alg}
	Problem (\ref{Problem_2}) is complicated because: i) it is a MOOP to derive a trade-off among the profit functions of all MNOs; ii) there are binary variables, $\boldsymbol{\rm{A}}$; and iii) the expected rate functions in the objective function and constraints C1 and C2 are nonconvex stochastic functions with no closed-form expression. We aim to propose an efficient algorithm with reasonable computational complexity to address problem (\ref{Problem_2}). To tackle these difficulties, first, we employ a combination of the $\epsilon$-constraint and weighted sum methods to transform the MOOP into a single objective form. Next, we deal with the binary variables that produce a disjoint feasible solution set that is an obstacle to solve problem (\ref{Problem_2}). Similar to the technique taken in \cite{Che2014}, we convert the binary variables into continuous ones by applying a penalty function. Finally, we use the constrained stochastic successive convex approximation (CSSCA) algorithm to approximate the nonconvex expected rate functions in the objective function and the constraints with some convex surrogate functions. By doing so, problem (\ref{Problem_2}) is transformed into a convex problem. Hence, in each iteration of the CSSCA algorithm, we solve a convex optimization problem by using off-the-shelf optimization toolboxes such as CVX \cite{CVX}, \cite{Boyd2004}. The details of the proposed method to address the optimization problem (\ref{Problem_2}) are explained in the following.
	\vspace{-0.9 em}
	\subsection{Transforming the MOOP (\ref{Problem_2}) into a Single Objective Optimization Problem}
	Problem (\ref{Problem_2}) is a multiobjective nonconvex optimization problem with various constraints. At the first stage, to solve problem (\ref{Problem_2}), we convert the MOOP (\ref{Problem_2}) into a single objective form. To this end, we use a combination of the weighted sum method and the $\epsilon$-constraint method, which are two basic generation methods widely used to solve MOOPs \cite{Miettinen1999}.		
	
	In the weighted sum method, we generate a single objective function for seller MNOs by linearly combining seller MNOs' profit function. Furthermore, in the weighted sum method, we incorporate a set of weight factors: $\{v_{s}, s\in\mathcal{S}\}$, where the sum of the weights is equal to one, i.e., $\sum_{s\in\mathcal{S}}{v}_{s}=1$. These already given weights indicate the relative importance of each seller MNO's profit function. Similarly, we use the weighted sum method to generate a single objective function for buyer MNOs by linearly combining buyer MNOs' profit function. In addition, we incorporate a set of weight factors: $\{{w}_{b}, b\in\mathcal{B}\}$, where the sum of the weights is equal to one, i.e., $\sum_{b\in\mathcal{B}}{w}_{b}=1$. These already given weights reflect the relative importance of each buyer MNO's profit function. After that, because the weighted profit function of seller MNOs and the weighted profit function of buyer MNOs are two conflicting objective functions, we employ the $\epsilon$-constraint method. Due to the fact that our main goals are to enhance spectrum utilization and increase the profit of MNOs, we keep the total profit function of buyer MNOs as the primary objective function and cast the total profit function of seller MNOs to the constraint set. Thus, by varying the value of $\epsilon$ and solving the corresponding optimization problem, we can adjust the lower bound for the profit of the seller MNOs and further obtain all Pareto optimal solutions. By doing so, the MOOP (\ref{Problem_2}) is transformed into a single objective optimization problem as
	\begin{equation} \label{Problem_3}
	\begin{aligned}
	&\max\limits_{\boldsymbol{\rm{A}},\boldsymbol{\rm{P}}}
	&\quad
	& \sum_{b\in\mathcal{B}}^{}w_{b}u_{b}(\boldsymbol{\rm{A}},\boldsymbol{\rm{P}})\\
	& \text{s.t.} &
	&\text{C0}:\sum_{s\in\mathcal{S}}^{}v_{s}u_{s}(\boldsymbol{\rm{A}},\boldsymbol{\rm{P}})\geq \epsilon,\qquad \text{C1 -- C6}.
	\end{aligned}
	\end{equation}
	According to constraint C0, the total profit function of seller MNOs should not be less than $\epsilon$. In the constraint, $\epsilon$ derives a trade-off between the profit of the seller MNOs and the buyer MNOs.
	
	Note that according to constraint C0 in problem (\ref{Problem_3}), the total profit of the seller MNOs, i.e., $\sum_{s\in\mathcal{S}}v_{s}u_{s}(\boldsymbol{\rm{A}},\boldsymbol{\rm{P}})$, is greater than or equal to $\epsilon$. To further clarify the significance of the parameter $\epsilon$ on the feasibility and solution of problem (\ref{Problem_3}), we consider three following cases:
	\begin{enumerate}
		\item{
			if $\epsilon$ = 0, then problem (\ref{Problem_3}) converts into the problem of maximizing the total profit of the buyer MNOs subject to constraints C1--C6. It can be solved as proposed in this paper.
		}
		\item {
			if $\epsilon = U^{\text{max}}$, where $U^{\text{max}}$ is the maximum profit of the seller MNOs, then problem (\ref{Problem_3})  converts into the problem of maximizing the total profit of the seller MNOs subject to constraints C1--C6. It can be solved as proposed in this paper.
		}
		\item {
			if $\epsilon\ge U^{\text{max}}$, then problem (\ref{Problem_3}) is infeasible.
		}
	\end{enumerate}
	
	Considering the above cases, it can be easily concluded that problem (\ref{Problem_3}) and its solution are very sensitive to the value of $\epsilon$. With this parameter, a trade-off between the total profit of the seller MNOs and the total profit of the buyer MNOs can be derived, leading to a change in the priority of the objective functions. Specifically, when the value chosen for $\epsilon$ is high, more emphasis is put on the maximization of the total profit of the seller MNOs, while lower values of $\epsilon$ assign a higher priority for the maximization of the total profit of the buyer MNOs. To find this specific value of $\epsilon$ when our goal is to maximize the total profit of the buyer MNOs, we propose the following algorithm.
	
	From cases (1) and (2) we derive that the maximum value that $\epsilon$ can take without making (\ref{Problem_3}) infeasible is $U^{\text{max}}$. Because $U^{\text{max}}$ is the maximum profit of the seller MNOs, we can obtain its value by solving the optimization problem of maximizing the profit of the seller MNOs as
	\begin{equation} \label{Problem_U}
	\begin{aligned}
	&\max\limits_{\boldsymbol{\rm{A}},\boldsymbol{\rm{P}}}
	&\quad
	&\sum_{s\in\mathcal{S}}^{}v_{s}u_{s}(\boldsymbol{\rm{A}},\boldsymbol{\rm{P}})\\
	& \text{s.t.}
	&& \text{C1 -- C6}.
	\end{aligned}
	\end{equation}
	By solving problem (\ref{Problem_U}), the maximum value of $\epsilon$, i.e., $U^{\text{max}}$ that avoids infeasibility of problem (\ref{Problem_3}) can be determined. It is worth mentioning that problem (\ref{Problem_U}) is as complex as problem (\ref{Problem_3}). Therefore, to address problem (\ref{Problem_U}), we can employ the algorithm proposed in Section \ref{App}.
	
	Because different values of $\epsilon$ lead to trade-offs between the total profit of the buyer MNOs and the total profit of the seller MNOs, we obtain a value for $\epsilon$ that maximizes the ratio of the total profit of the buyer MNOs to the total profit of the seller MNOs. Thus, $\epsilon$ is expressed as
	\begin{equation} \label{phi}
	\begin{aligned}
	\epsilon = \phi\:U^{\text{max}},
	\end{aligned}
	\end{equation}
	where $\phi$ is a positive  value in the range of $[0,1]$, and $U^{\text{max}}$ is the maximum total profit of the seller MNOs obtained by solving problem (\ref{Problem_U}). The ratio between the total profit of the seller MNOs and the total profit of the buyer MNOs varies depending on the value of $\phi$. However, for a specific $\phi$, this ratio reaches a maximum value.	To find this point, we test different values of $\phi$ in which the maximum total profit of the buyer MNOs to the total profit of the seller MNOs ratio is achieved.
	\subsection{Tackling the Binary Variables in Problem (\ref{Problem_3})}
	Although problem (\ref{Problem_3}) is a single objective function, the binary constraint C6 makes it complicated to solve. To handle this difficulty, similar to the approach in \cite{Che2014}, we replace the binary constraint C6 with the following equivalent constraints:
	\begin{equation}
	\begin{aligned}
	&\text{C6.1}:\sum_{s\in\mathcal{S}}^{}\sum_{b\in\mathcal{B}}^{}\sum_{l_{s}\in\mathcal{L}_{s}}^{} \left(a_{l_{s},b} - {\left(a_{l_{s},b}\right)}^{2}\right) \leq 0,\\
	&\text{C6.2}:0 \leq a_{l_{s},b}\leq 1,
	\forall s\in \mathcal{S}, \forall b\in \mathcal{B}, \forall l_{s} \in \mathcal{L}_{s}.
	\end{aligned}
	\end{equation}
	
	 Note that, with respect to constraint C6.2, the sub-band sharing variable, i.e., $a_{l_{s},b}$, is a continuous value between zero and one, and with respect to constraint C6.1, $a_{l_{s},b}$ is forced to be either one or zero because $(a_{l_{s},b} - {(a_{l_{s},b})}^{2})$ always takes nonnegative values but constraint C6.1 pushes $(a_{l_{s},b} - {(a_{l_{s},b})}^{2})$ to be nonpositive.
	
	Instead of constraint C6, we add these new constraints to problem (\ref{Problem_3}) and equivalently rewrite it as
	\begin{equation}\label{Problem_31}
	\begin{aligned}
	&\max\limits_{\boldsymbol{\rm{A}},\boldsymbol{\rm{P}}}
	&\quad
	& \sum_{b\in\mathcal{B}}^{}w_{b}u_{b}(\boldsymbol{\rm{A}},\boldsymbol{\rm{P}})\\
	& \text{s.t.} &
	&\text{C0 -- C5}, \text{C6.1, C6.2}.
	\end{aligned}
	\end{equation}
	
	Although all variables in problem (\ref{Problem_31}) are continuous, we eventually aim to find a binary value for the	$a_{l_{s},b}$ variable. To reach this goal, we reformulate problem (\ref{Problem_31}) as
	\begin{equation}
	\begin{aligned} \label{Problem_5}
	&\max\limits_{\boldsymbol{\rm{A}},\boldsymbol{\rm{P}}}
	\quad
	\sum_{b\in\mathcal{B}}^{}w_{b}u_{b}\left(\boldsymbol{\rm{A}},\boldsymbol{\rm{P}}\right)-
	\vartheta\sum_{s\in\mathcal{S}}^{}\sum_{b\in\mathcal{B}}^{}\sum_{l_{s}\in\mathcal{L}_{s}}^{}\left(a_{l_{s},b} - {\left(a_{l_{s},b}\right)}^{2}\right)\\
	& \text{s.t.}
	\quad \quad\rm{C0 - C5}, \rm{C6.2},
	\end{aligned}
	\end{equation}
	where $\vartheta$ is constant and acts as a penalty factor to penalize the	violation of constraint C6.1. It can be proven that for sufficiently large values of $\vartheta$, the optimization problem (\ref{Problem_5}) is equivalent to (\ref{Problem_31}).		
	\begin{lem}\label{thm_4}
		For sufficiently large values of $\vartheta$, problem (\ref{Problem_5}) is equivalent to problem (\ref{Problem_31}) \cite{Che2014}, \cite{Aslani2019}.
	\end{lem}
	\begin{proof}
		By using the abstract Lagrangian duality \cite{Che2014}, the primal problem of (\ref{Problem_31}) can be expressed as
		\begin{equation}\label{pimal}
		\begin{aligned}
		p^* =\displaystyle \max_{\substack{(\boldsymbol{\rm{A}},\boldsymbol{\rm{P}}})\in\mathcal{F}}\displaystyle \min_{\substack{\nu_1\geq 0}}\mathcal{L}(\boldsymbol{\rm{A}},\boldsymbol{\rm{P}},\vartheta),
		\end{aligned}
		\end{equation} 
		where $\displaystyle\mathcal{L}(\boldsymbol{\rm{A}},\boldsymbol{\rm{P}},\vartheta) =\sum_{b\in\mathcal{B}}w_{b}u_{b}(\boldsymbol{\rm{A}},\boldsymbol{\rm{P}})- \vartheta \sum_{s\in\mathcal{S}}\sum_{b \in \mathcal{B}}\sum_{l_{s}\in \mathcal{L}_{s}}(a_{l_{s},b}-a_{l_{s},b}^2)$, $\mathcal{F}$ denotes the feasible set spanned by constraints C0--C5, and $p^*$ is the optimal solution of the primal problem (\ref{pimal}). On the other hand, the dual problem of (\ref{Problem_31}) is
		\begin{equation}\label{duality}
		\begin{aligned}
		\displaystyle d^* = \displaystyle\min_{\substack{\vartheta\geq 0}}\max_{\substack{(\boldsymbol{\rm{A}},\boldsymbol{\rm{P}})}\in\mathcal{F}}\displaystyle \mathcal{L}(\boldsymbol{\rm{A}},\boldsymbol{\rm{P}},\vartheta)\overset{\triangle}{=}\displaystyle\min_{\substack{\vartheta\geq 0}} \Phi(\vartheta),
		\end{aligned}
		\end{equation}
		where $ \Phi(\vartheta)=\displaystyle \max_{\substack{(\boldsymbol{\rm{A}},\boldsymbol{\rm{P}})\in\mathcal{F}}}\displaystyle \mathcal{L}(\boldsymbol{\rm{A}},\boldsymbol{\rm{P}},\vartheta) $ and $ d^* $ is the optimal solution of the dual problem (\ref{duality}). Based on the weak duality \cite{Boyd2004}, the following inequality is met by the primal problem (\ref{pimal}) and the dual problem (\ref{duality}), i.e.,
		\begin{equation}\label{gap}
		\begin{aligned}	
		&p^*=\max_{\substack{(\boldsymbol{\rm{A}},\boldsymbol{\rm{P}})\in\mathcal{F}}}\min_{\substack{\vartheta\geq 0}}\displaystyle \mathcal{L}(\boldsymbol{\rm{A}},\boldsymbol{\rm{P}},\vartheta)\leq \displaystyle \min_{\substack{\vartheta\geq 0}}\max_{\substack{(\boldsymbol{\rm{A}},\boldsymbol{\rm{P}})\in\mathcal{F}}}\displaystyle \mathcal{L}(\boldsymbol{\rm{A}},\boldsymbol{\rm{P}},\vartheta)\\
		&=\min_{\substack{\vartheta\geq 0}} \Phi(\vartheta)=d^*.
		\end{aligned}
		\end{equation}
		
		We note that $ \mathcal{L}(\boldsymbol{\rm{A}},\boldsymbol{\rm{P}},\vartheta) $ is a decreasing function in $ \vartheta$ because $\sum_{s\in\mathcal{S}}\sum_{b \in \mathcal{B}}\sum_{l_{s}\in \mathcal{L}_{s}}(a_{l_{s},b}-a_{l_{s},b}^2)\geq 0$. Thus, $\Phi(\vartheta)$ decreases with $ \vartheta $ and is bounded from below by the optimal solution $p^*$ based on (\ref{gap}). Assuming $\vartheta^* $, $\boldsymbol{\rm{A}}^*$, and $\boldsymbol{\rm{P}}^{*}$ as the optimal point of the dual problem (\ref{duality}), we study two following cases.
		
		For the first case, we assume that $ \sum_{s\in\mathcal{S}}\sum_{b \in \mathcal{B}}\sum_{l_{s}\in \mathcal{L}_{s}}(a_{l_{s},b}-a_{l_{s},b}^2)= 0 $. Therefore, the optimal point of the dual problem (\ref{duality}), i.e.,  $ \vartheta^* $, $\boldsymbol{\rm{A}}^*$, and $\boldsymbol{\rm{P}}^{*}$, is further a feasible solution to  problem (\ref{Problem_31}). Consequently, by substituting $\boldsymbol{\rm{A}}^*$ and $\boldsymbol{\rm{P}}^{*}$ into problem (\ref{Problem_31}), we have
		\begin{equation}\label{case-1}
		\begin{aligned}
		p^*&=\max_{\substack{(\boldsymbol{\rm{A}}^{*},\boldsymbol{\rm{P}}^{*})\in\mathcal{F}^{*} }}\min_{\substack{\vartheta^*\geq 0}}\displaystyle \mathcal{L}(\boldsymbol{\rm{A}}^{*},\boldsymbol{\rm{P}}^{*},\vartheta^*)\geq \sum_{b\in\mathcal{B}}w_{b}u_{b}(\boldsymbol{\rm{A}}^{*},\boldsymbol{\rm{P}}^{*})\\
		&=\mathcal{L}(\boldsymbol{\rm{A}}^{*},\boldsymbol{\rm{P}}^{*},\vartheta^*)= \Phi(\vartheta^*)=\displaystyle \min_{\substack{\vartheta\geq 0}} \Phi(\vartheta)=d^*.
		\end{aligned}
		\end{equation}
		By combining \eqref{gap} and \eqref{case-1}, we can conclude that the gap between the primal problem (\ref{pimal}) and the dual problem (\ref{duality}) is zero, i.e.,
		\begin{equation}\label{gap-zero}
		\begin{aligned}	
		p^*&=\max_{\substack{(\boldsymbol{\rm{A}},\boldsymbol{\rm{P}})\in\mathcal{F} }}\min_{\substack{\vartheta\geq 0}}\displaystyle \mathcal{L}(\boldsymbol{\rm{A}},\boldsymbol{\rm{P}},\vartheta)= \displaystyle \min_{\substack{\vartheta\geq 0}}\displaystyle \max_{\substack{(\boldsymbol{\rm{A}},\boldsymbol{\rm{P}})\in\mathcal{F} }}\mathcal{L}(\boldsymbol{\rm{A}},\boldsymbol{\rm{P}},\vartheta)\\
		&=d^*.
		\end{aligned}
		\end{equation}
		In addition, with respect to $ \vartheta $, the monotonically decreasing function of $ \Phi({\vartheta}) $ implies that we have $ \Phi({\vartheta})=d^*, ~\forall \vartheta \geq {\vartheta^*} $. This means that the optimal solution of problem (\ref{Problem_5}) is equal to the optimal solution of problem (\ref{Problem_31}).
		
		For the second case, we consider that $ \sum_{s\in\mathcal{S}}\sum_{b \in \mathcal{B}}\sum_{l_{s}\in \mathcal{L}_{s}}(a_{l_{s},b}-a_{l_{s},b}^2)> 0$. Since $ \Phi({\vartheta}) $ is a monotonically decreasing function with respect to $ \vartheta$, dual problem (\ref{duality}) tends to $-\infty $ and it is unbounded from below. This result states a contradiction with (\ref{gap}) where dual problem (\ref{duality}) is finite and positive, and it is bounded from below by the optimal solution of primal problem (\ref{pimal}), i.e., $p^*$. Therefore, the first case, i.e., $ \sum_{s\in\mathcal{S}}\sum_{b \in \mathcal{B}}\sum_{l_{s}\in \mathcal{L}_{s}}(a_{l_{s},b}-a_{l_{s},b}^2)= 0$, holds for the optimal solution and the proof is completed.
	\end{proof}
	\subsection{CSSCA-Based Algorithm to Solve Problem (\ref{Problem_5})}\label{App}
	Although the binary variables were transformed into continuous ones in problem (\ref{Problem_5}), it is still difficult to solve. The reason is that the mathematical expectations in (\ref{r_m_}) and (\ref{r_n_}) are defined for the functions with the structure of $\log(1+x/y)$, and thus, (\ref{r_m_}) and (\ref{r_n_}) are stochastic nonconvex functions and have no closed-form expression, which leads to nonconvexity of the objective function and constraints C0--C2 with respect to $\boldsymbol{\rm{A}}$ and $\boldsymbol{\rm{P}}$.
	
	To address these difficulties in problem (\ref{Problem_5}), we apply the CSSCA method \cite{Liu2019} --\cite{Liu2018} to approximate nonconvex stochastic objective and constraint functions with convex surrogate functions to convert problem (\ref{Problem_5}) into a convex form that can be solved efficiently. This algorithm is based on solving a sequence of convex approximation problems. The output of the previously solved convex approximation problems is used to update optimization variables, which are the input to the approximation formulas employed to construct the next convex approximation problem. Accordingly, at each iteration, we successively improve our approximation of the original problem as convex by the output of the previous iterations.
	
	To solve problem (\ref{Problem_5}) by employing the CSSCA algorithm, we first restate problem (\ref{Problem_5}) in the general form of a nonconvex constrained stochastic optimization problem as
	\begin{equation} \label{Problem_7}
	\begin{aligned}
	&
	\min\limits_{(\boldsymbol{\rm{A}},\boldsymbol{\rm{P}})\in\mathcal{X}}
	- \sum_{b\in\mathcal{B}}^{}w_{b}\Big(\delta DN_{b} \bar{r}_{b}(\boldsymbol{\rm{A}},\boldsymbol{\rm{P}}) -c_{b}\left(\boldsymbol{\rm{A}}\right)\Big) \\
	&\qquad\qquad+ \vartheta\sum_{s\in\mathcal{S}}^{}\sum_{l_{s}\in\mathcal{L}_{s}}^{}\sum_{b\in\mathcal{B}}^{}\left(a_{l_{s},b} - {\left(a_{l_{s},b}\right)}^{2}\right)\\
	& \text{s.t.}\qquad
	\overline{\text{C}}0:\epsilon-\sum_{s\in\mathcal{S}}^{}v_{s}\Big(\delta DN_{s} \bar{r}_{s}(\boldsymbol{\rm{A}},\boldsymbol{\rm{P}})+\nu_{s}^{2}(\boldsymbol{\rm{A}}) -c_{s}\Big) \leq 0,\\
	&\qquad~~~~\overline{\text{C}}1:r^{\text{th}} - {\overline{r}}_{k}(\boldsymbol{\rm{A}},\boldsymbol{\rm{P}})\leq 0, 
	\qquad\qquad\qquad~~~\forall k\in\mathcal{O}, \\
	&\qquad~~~~\overline{\text{C}}2: -\Big(\delta DN_{b} \bar{r}_{b}(\boldsymbol{\rm{A}},\boldsymbol{\rm{P}}) -c_{b}(\boldsymbol{\rm{A}})\Big) \leq 0, 
	\qquad~\forall b\in\mathcal{B},
	\end{aligned}
	\end{equation}
	where $\mathcal{X}$ denotes the feasible set spanned by constraints C3, C4, C5, and C6.2, which is convex and deterministic. Note that $\overline{{r}}_{k}(\boldsymbol{\rm{A}},\boldsymbol{\rm{P}})=\mathbb{E}\left[r_{k}(\boldsymbol{\rm{A}},\boldsymbol{\rm{P}};\xi)\right],\forall k\in\mathcal{O}$ is a nonconvex stochastic function, in which the expectation is over the random vector $\xi$, and $\xi$ denotes the random vector consisting of channel fading samples, transmit power samples, and the number and location of users and BS samples. Moreover, the total number of nonconvex stochastic constraints is $2B+S+1$.
	\begin{algorithm}[!t]
		\SetKwFunction{Range}{range}
		\SetKw{KwTo}{in}\SetKwFor{For}{for}{\string:}{}%
		\SetKwIF{If}{ElseIf}{Else}{if}{:}{elif}{else:}{}%
		\SetAlgoNoEnd
		\SetAlgoNoLine%
		\SetKwInOut{Input}{Input}
		\SetKwInOut{Output}{Output}
		\BlankLine
		\textbf{Initialization: }{maximum number of iterations (samples) $T$, $\vartheta>\!\!>1$, $t=0$, $(\boldsymbol{\rm{A}}^{0},\boldsymbol{\rm{P}}^{0})\in\mathcal{X}$, $\tau_{i}, \forall i$, $\{\rho^{t}\}_{t=0}^{T}$, $\{\beta^{t}\}_{t=0}^{T}$}.\\
		\Indm
		\Indp
		\textbf{repeat} \\
		\Indp
		Obtain a sample of $\xi^{t}$ at iteration $t$.\\
		Given $\rho^{t}$, update the surrogate functions
		$\bar{f}_{0}^{t}(\boldsymbol{\rm{A}},\boldsymbol{\rm{P}})$, $\bar{f}_{1}^{t}(\boldsymbol{\rm{A}},\boldsymbol{\rm{P}})$, $\bar{f}_{k+1}^{t}(\boldsymbol{\rm{A}},\boldsymbol{\rm{P}}),\forall k\in\mathcal{O}$, and $\bar{f}_{b+|\mathcal{O}|+1}^{t}(\boldsymbol{\rm{A}},\boldsymbol{\rm{P}}),\forall b\in\mathcal{B}$, having $\xi^t$, $\boldsymbol{\rm{A}}^{t}$, and $\boldsymbol{\rm{P}}^{t}$.\\
		Solve problem (\ref{Problem_7_1}) to get $\bar{\boldsymbol{\rm{A}}}^t$ and $\bar{\boldsymbol{\rm{P}}}^t$.\\
		\If{\rm{problem (\ref{Problem_7_1}) is infeasible}}{Solve problem (\ref{Problem_8}) to get $\bar{\boldsymbol{\rm{A}}}^t$ and $\bar{\boldsymbol{\rm{P}}}^t$.}
		\textbf{end  if.} \\
		Given $\beta^{t}$, update $\boldsymbol{\rm{A}}^{t+1}$ and $\boldsymbol{\rm{P}}^{t+1}$ according to (\ref{update_new_1}) and (\ref{update_new_2}), respectively.\\
		Set $t = t + 1$.\\
		\Indm
		\textbf{until} $t = T$.
		\caption{Proposed CSSCA-based spectrum sharing algorithm\label{SHA}}
	\end{algorithm}
	\begin{figure}[!t]
		\centering
		\begin{tikzpicture}[node distance = 1 cm]
		\node [b1]  (box1)
		{Using $\xi^{t}$, $\boldsymbol{\rm{A}}^{t}$, and $\boldsymbol{\rm{P}}^{t}$, update all surrogate functions};
		
		\node [b2, below = 0.7cm of box1]  (box2)
		{Solve problem (\ref{Problem_7_1}) using the CVX toolbox};
		
		\node [c1, below = 0.7cm of box2]  (box3)
		{If problem (\ref{Problem_7_1}) is infeasible};
		
		\node [b2, right = 0cm of box3, xshift=25pt]  (box4)
		{Solve problem (\ref{Problem_8}) using the CVX toolbox};
		
		\node [coordinate, below =2.5cm of box4] (text1) {};
		\node [b1, below = 0.7cm of box3]  (box5)
		{Update $\boldsymbol{\rm{A}}^{t+1}$ and $\boldsymbol{\rm{P}}^{t+1}$ using (\ref{update_new_1}) and (\ref{update_new_2}), respectively};
		
		\node [coordinate, left = 0cm of box5,xshift=-7.5pt] (text2) {};
		\node [coordinate, above = 0cm of text2,yshift=204pt] (text3) {};
		\path [line] (box1) -- (box2) node[draw=none,font=\footnotesize,fill=none,midway,above] {};
		\path [line] (box2) -- (box3) node[draw=none,font=\footnotesize,fill=none,midway] {$\bar{\boldsymbol{\rm{A}}}^{t}$ and $\bar{\boldsymbol{\rm{P}}}^{t}$};
		\path [line] (box3) -- (box4) node[draw=none,font=\footnotesize,fill=none,midway,above] {Yes};
		\path [line] (box3) -- (box5) node[draw=none,font=\footnotesize,fill=none,midway,left]{No};
		\path [line] (box4) -- (text1) -- (box5) node[draw=none,font=\footnotesize,fill=none,midway,above] {$\bar{\boldsymbol{\rm{A}}}^{t}$ and $\bar{\boldsymbol{\rm{P}}}^{t}$};
		\path [line] (box5) -- (text2) -- (text3)  node[draw=none,rotate=90,font=\footnotesize,fill=none,midway,above] {$\boldsymbol{\rm{A}}^{t+1}$ and $\boldsymbol{\rm{P}}^{t+1}$} -- (box1);
		
		\end{tikzpicture}
		\caption{Block diagram of the proposed CSSCA-based spectrum sharing algorithm.}
		\label{BD}	
	\end{figure}
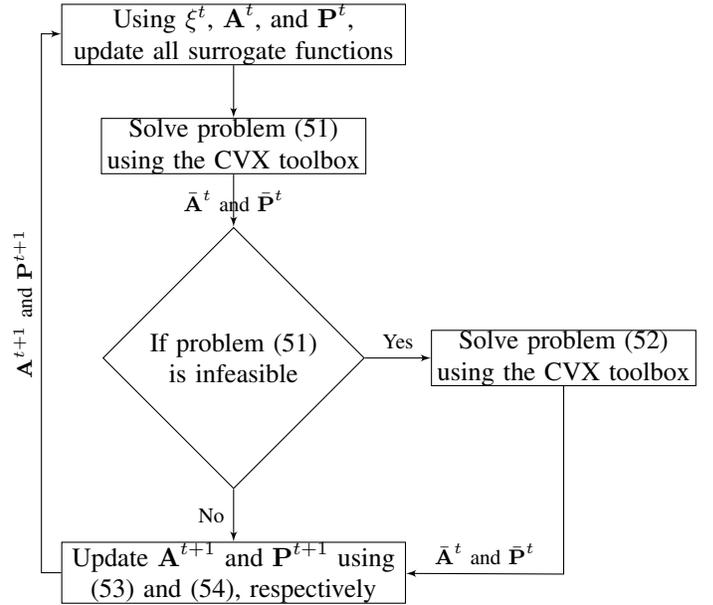

	Now, we apply the CSSCA algorithm, which is an iterative algorithm in nature, resulting in an efficient solution to problem (\ref{Problem_7}). The proposed CSSCA-based spectrum sharing algorithm is summarized in Algorithm \ref{SHA}, and its block diagram is shown in Fig. \ref{BD}. The CSSCA algorithm in each iteration consists of three steps, which are explained in the following to solve problem (\ref{Problem_7}).
	
	{\bf Step 1:} Given the approximation center $(\boldsymbol{\rm{A}}^{t},\boldsymbol{\rm{P}}^{t})$ and the current sample $\xi^{t}$ at the iteration $t$, the surrogate function of the objective function and constraints $\overline{\text{C}}0$--$\overline{\text{C}}2$ in problem (\ref{Problem_7}), denoted by $\{\bar{f}_{i}(\boldsymbol{\rm{A}},\boldsymbol{\rm{P}}), \forall i\in 0 \cup 1\cup \mathcal{O}\cup\mathcal{B}\}$, are updated. It is worth mentioning that $\bar{f}_{0}(\boldsymbol{\rm{A}},\boldsymbol{\rm{P}})$ is the surrogate function of the objective function in problem (\ref{Problem_7}), i.e.,  $f_{0}(\boldsymbol{\rm{A}},\boldsymbol{\rm{P}}) \overset{\Delta}{=} -\sum_{b\in\mathcal{B}}^{}w_{b}(\delta DN_{b} \bar{r}_{b}(\boldsymbol{\rm{A}},\boldsymbol{\rm{P}}) -c_{b}(\boldsymbol{\rm{A}}))+ \vartheta\sum_{s\in\mathcal{S}}^{}\sum_{l_{s}\in\mathcal{L}_{s}}^{}\sum_{b\in\mathcal{B}}^{}\left(a_{l_{s},b} - (a_{l_{s},b})^{2}\right)$, while $\{\bar{f}_{1}(\boldsymbol{\rm{A}},\boldsymbol{\rm{P}})\}$, $\{\bar{f}_{k}(\boldsymbol{\rm{A}},\boldsymbol{\rm{P}}), \forall k\in\mathcal{O}\}$, and $\{\bar{f}_{b}(\boldsymbol{\rm{A}},\boldsymbol{\rm{P}}), \forall b\in\mathcal{B}\}$ are the surrogate functions for the constraint functions $\overline{\text{C}}0$, i.e., $\{f_{1}(\boldsymbol{\rm{A}},\boldsymbol{\rm{P}}) \overset{\Delta}{=}\epsilon-\sum_{s\in\mathcal{S}}^{}v_{s}\Big(\delta DN_{s} \bar{r}_{s}(\boldsymbol{\rm{A}},\boldsymbol{\rm{P}})+\nu_{s}^{2}(\boldsymbol{\rm{A}}) -c_{s}\Big) \leq 0\}$, $\overline{\text{C}}1$, i.e., $\{f_{k}(\boldsymbol{\rm{A}},\boldsymbol{\rm{P}}) \overset{\Delta}{=}r^{\text{th}} - {\overline{r}}_{k}(\boldsymbol{\rm{A}},\boldsymbol{\rm{P}})\leq 0,\forall k\in\mathcal{O}\}$, and $\overline{\text{C}}2$, i.e., $\{f_{b}(\boldsymbol{\rm{A}},\boldsymbol{\rm{P}}) \overset{\Delta}{=}-\Big(\delta DN_{b} \bar{r}_{b}(\boldsymbol{\rm{A}},\boldsymbol{\rm{P}}) -c_{b}(\boldsymbol{\rm{A}})\Big) \leq 0, \forall b\in\mathcal{B}\}$ in problem (\ref{Problem_7}), respectively. Generally, $\{\bar{f}_{i}(\boldsymbol{\rm{A}},\boldsymbol{\rm{P}}), \forall i\in 0 \cup 1\cup \mathcal{O}\cup\mathcal{B}\}$ can be updated by
	\begin{equation}\label{sur_11}
	\begin{aligned}
	&\bar{f}_{i}^{t}(\boldsymbol{\rm{A}},\boldsymbol{\rm{P}}) = {f}^{t}_{i}+ (\mathbf{f}_{\boldsymbol{\rm{A}},i}^{t})^{T}(\boldsymbol{\rm{A}}-\boldsymbol{\rm{A}}^{t}) +(\mathbf{f}_{\boldsymbol{\rm{P}},i}^{t})^{T}(\boldsymbol{\rm{P}}-\boldsymbol{\rm{P}}^{t})\\
	&+\tau_{i}\left({||\boldsymbol{\rm{A}}-\boldsymbol{\rm{A}}^{t}||}^{2}+ {||\boldsymbol{\rm{P}}-\boldsymbol{\rm{P}}^{t}||}^{2}\right),\forall i\in 0 \cup 1\cup \mathcal{O}\cup\mathcal{B},
	\end{aligned}
	\end{equation}
	in which $\tau_{i}>0$ is a constant value, the constant $f_{i}^{t}$ is an approximation for $f_{i}(\boldsymbol{\rm{A}}^{t},\boldsymbol{\rm{P}}^{t})$, $\mathbf{f}_{\boldsymbol{\rm{A}},i}^{t}$ and $\mathbf{f}_{\boldsymbol{\rm{P}},i}^{t}$ are approximations of the partial derivatives $\nabla_{\boldsymbol{\rm{A}}} f_{i}(\boldsymbol{\rm{A}}^{t},\boldsymbol{\rm{P}}^{t})$ and $\nabla_{\boldsymbol{\rm{P}}} f_{i}(\boldsymbol{\rm{A}}^{t},\boldsymbol{\rm{P}}^{t})$ with respect to $\boldsymbol{\rm{A}}$ and $\boldsymbol{\rm{P}}$, respectively, and $\nabla_{\boldsymbol{\rm{A}}}$ and $\nabla_{\boldsymbol{\rm{P}}}$ are the gradient operations with respect to $\boldsymbol{\rm{A}}$ and $\boldsymbol{\rm{P}}$, respectively. 
	
	It is noteworthy that the convex surrogate function $\bar{f}_{i}^{t}(\boldsymbol{\rm{A}},\boldsymbol{\rm{P}})$ in (\ref{sur_11}) can be regarded as a convex approximation of $f_{i}(\boldsymbol{\rm{A}},\boldsymbol{\rm{P}})$ around the point $(\boldsymbol{\rm{A}}^{t},\boldsymbol{\rm{P}}^t)$ at the iteration $t$ so that the first term in (\ref{sur_11}) indicates the sample average approximation of $f_{i}(\boldsymbol{\rm{A}},\boldsymbol{\rm{P}})$ at the iteration $t$. The second and third terms in (\ref{sur_11}) relate to the incremental estimate of $\nabla_{\boldsymbol{\rm{A}}} f_{i}(\boldsymbol{\rm{A}}^{t},\boldsymbol{\rm{P}}^{t})$ and $\nabla_{\boldsymbol{\rm{P}}} f_{i}(\boldsymbol{\rm{A}}^{t},\boldsymbol{\rm{P}}^{t})$, respectively, which are not available. In other words, the second and third terms in (\ref{sur_11}) estimate the unknown $\nabla_{\boldsymbol{\rm{A}}} f_{i}(\boldsymbol{\rm{A}}^{t},\boldsymbol{\rm{P}}^{t})$ and $\nabla_{\boldsymbol{\rm{P}}} f_{i}(\boldsymbol{\rm{A}}^{t},\boldsymbol{\rm{P}}^{t})$, respectively, by their realizations collected over the iterations so that this estimation is expected to become more and more accurate as $t$ increases. Finally, the last quadratic term of (\ref{sur_11}) guarantees strong convexity.
	
	For the given sample $\xi^{t}$, we can update $f_{0}^{t}$,  $f_{1}^{t}$, $\{f_{k}^{t}, \forall k\in\mathcal{O}\}$, and $\{f_{b}^{t}, \forall b\in\mathcal{B}\}$, respectively, by (\ref{F_1}), (\ref{F_2}), (\ref{F_3}), and (\ref{F_4}) at the top of next page, where $f_{i}^{-1}=0, \forall i\in 0\cup 1\cup\mathcal{O}\cup\mathcal{B}$ and $\rho^{t}\in(0,1]$ is a sequence to be properly chosen.
	\begin{figure*}[!t]
	\begin{flushleft}
		\begin{equation}\label{F_1}
		\begin{aligned}
		f^{t}_{0}=(1-\rho^{t})f^{t-1}_{0}+\rho^{t}\left(-\sum_{b\in\mathcal{B}}^{}w_{b}\left(\delta DN_{b} r_{b}(\boldsymbol{\rm{A}}^{t},\boldsymbol{\rm{P}}^{t};\xi^{t}) -c_{b}(\boldsymbol{\rm{A}}^{t})\right)+\vartheta\sum_{s\in\mathcal{S}}^{}\sum_{l_{s}\in\mathcal{L}_{s}}^{}\sum_{b\in\mathcal{B}}^{}\left(a_{l_{s},b}^{t} - (a_{l_{s},b}^{t})^{2}\right)\right).
		\end{aligned}
		\end{equation}
		\begin{equation} \label{F_2}
		\begin{aligned}
		f^{t}_{1}=(1-\rho^{t})f^{t-1}_{1}+\rho^{t}\left(\epsilon-\sum_{s\in\mathcal{S}}^{}v_{s}\left(\delta DN_{s} r_{s}(\boldsymbol{\rm{A}}^{t},\boldsymbol{\rm{P}}^{t};\xi^{t}) +\nu_{s}^{2}(\boldsymbol{\rm{A}}^{t})-c_{s}\right)\right).
		\end{aligned}
		\end{equation}
		\begin{equation}\label{F_3}
		\begin{aligned}
		f^{t}_{k}&=(1-\rho^{t})f^{t-1}_{k}+\rho^{t}\left(r^{\text{th}}- r_{k}(\boldsymbol{\rm{A}}^{t},\boldsymbol{\rm{P}}^{t};\xi^{t})\right),\qquad\qquad\forall k\in\mathcal{O}.
		\end{aligned}
		\end{equation}
		\begin{equation}\label{F_4}
		\begin{aligned}
		f^{t}_{b}&=(1-\rho^{t})f^{t-1}_{b}-\rho^{t}\left(\delta DN_{b} r_{b}(\boldsymbol{\rm{A}}^{t},\boldsymbol{\rm{P}}^{t};\xi^{t}) -c_{b}(\boldsymbol{\rm{A}}^{t})\right),\qquad\qquad\forall b\in\mathcal{B}.
		\end{aligned}
		\end{equation}
	\end{flushleft}
	\hrulefill
	\vspace*{4pt}
	\end{figure*}
	
	Further, $\mathbf{f}_{\boldsymbol{\rm{A}},i}^{t}$ and $\mathbf{f}_{\boldsymbol{\rm{P}},i}^{t}$, for all $i\in 0\cup 1\cup\mathcal{O}\cup\mathcal{B}$ are updated recursively according to
	\begin{equation}
	\begin{aligned} \label{u_0A}
	&\mathbf{f}_{\boldsymbol{\rm{A}},0}^{t}=(1-\rho^{t})\mathbf{f}_{\boldsymbol{\rm{A}},0}^{t-1}+\rho^{t}(-\sum_{b\in\mathcal{B}}^{}w_{b}(\delta DN_{b} \nabla_{\boldsymbol{\rm{A}}}r_{b}(\boldsymbol{\rm{A}}^{t},\boldsymbol{\rm{P}}^{t};\xi^{t}) -\\
	&\sum_{s\in\mathcal{S}}^{}\sum_{l_{s}\in\mathcal{L}_{s}}^{}\varphi_{s,b})+\vartheta\sum_{s\in\mathcal{S}}^{}\sum_{l_{s}\in\mathcal{L}_{s}}^{}\sum_{b\in\mathcal{B}}^{}\left(1 - 2(a_{l_{s},b}^{t})\right)),
	\end{aligned}
	\end{equation}
	\begin{equation}
	\begin{aligned} \label{u_0P}
	\mathbf{f}_{\boldsymbol{\rm{P}},0}^{t}=(1-\rho^{t})\mathbf{f}_{\boldsymbol{\rm{P}},0}^{t-1}+\rho^{t}(-\sum_{b\in\mathcal{B}}^{}w_{b}\delta DN_{b} \nabla_{\boldsymbol{\rm{P}}}r_{b}(\boldsymbol{\rm{A}}^{t},\boldsymbol{\rm{P}}^{t};\xi^{t})),
	\end{aligned}
	\end{equation}
	\begin{equation}
	\begin{aligned} \label{u_1A}
	&\mathbf{f}_{\boldsymbol{\rm{A}},1}^{t}=(1-\rho^{t})\mathbf{f}_{\boldsymbol{\text{A}},1}^{t-1}+\rho^{t}\big(-\sum_{s\in\mathcal{S}}^{}v_{s}(\delta DN_{s} \nabla_{\boldsymbol{\text{A}}}r_{s}(\boldsymbol{\text{A}}^{t},\boldsymbol{\text{P}}^{t};\xi^{t}) + \\
	&\varphi_{s,b})\big),
	\end{aligned}
	\end{equation}
	\begin{equation}
	\begin{aligned} \label{u_1P}
	\mathbf{f}_{\boldsymbol{\text{P}},1}^{t}=(1-\rho^{t})\mathbf{f}_{\boldsymbol{\text{P}},1}^{t-1}+\rho^{t}(-\sum_{s\in\mathcal{S}}^{}v_{s}\left(\delta DN_{s} \nabla_{\boldsymbol{\rm{P}}}r_{s}(\boldsymbol{\rm{A}}^{t},\boldsymbol{\rm{P}}^{t};\xi^{t})\right)),
	\end{aligned}
	\end{equation}
	\begin{equation}
	\begin{aligned} \label{u_kA}
	\mathbf{f}_{\boldsymbol{\rm{A}},k}^{t}=(1-\rho^{t})\mathbf{f}_{\boldsymbol{\rm{A}},k}^{t-1}+\rho^{t}(- \nabla_{\boldsymbol{\rm{A}}}r_{k}(\boldsymbol{\rm{A}}^{t},\boldsymbol{\rm{P}}^{t};\xi^{t})), \forall k\in \mathcal{O},
	\end{aligned}
	\end{equation}
	\begin{equation}
	\begin{aligned} \label{u_kP}
	\mathbf{f}_{\boldsymbol{\rm{P}},k}^{t}=(1-\rho^{t})\mathbf{f}_{\boldsymbol{\rm{P}},k}^{t-1}+\rho^{t}(- \nabla_{\boldsymbol{\rm{P}}}r_{k}(\boldsymbol{\rm{A}}^{t},\boldsymbol{\rm{P}}^{t};\xi^{t})), \forall k\in \mathcal{O},
	\end{aligned}
	\end{equation}
	\begin{equation}
	\begin{aligned} \label{u_bA}
	&\mathbf{f}_{\boldsymbol{\rm{A}},b}^{t}=(1-\rho^{t})\mathbf{f}_{\boldsymbol{\rm{A}},b}^{t-1}+\rho^{t} \left(\delta DN_{b} \nabla_{\boldsymbol{\rm{A}}}r_{b}(\boldsymbol{\rm{A}}^{t},\boldsymbol{\rm{P}}^{t};\xi^{t}) -\varphi_{s,b}\right),\\
	&\forall b\in \mathcal{B},
	\end{aligned}
	\end{equation}
	\begin{equation}
	\begin{aligned} \label{u_bP}
	\mathbf{f}_{\boldsymbol{\rm{P}},b}^{t}=(1-\rho^{t})\mathbf{f}_{\boldsymbol{\rm{P}},b}^{t-1}+\rho^{t}\left(\delta DN_{b} \nabla_{\boldsymbol{\rm{P}}}r_{b}(\boldsymbol{\rm{A}}^{t},\boldsymbol{\rm{P}}^{t};\xi^{t})\right), \forall b\in \mathcal{B},
	\end{aligned}
	\end{equation}
	with $\mathbf{f}_{\boldsymbol{\rm{A}},i}^{-1}=\boldsymbol{0}$, $\mathbf{f}_{\boldsymbol{\rm{P}},i}^{-1}=\boldsymbol{0}$, $\nabla_{\boldsymbol{\rm{A}}}r_{k}(\boldsymbol{\rm{A}}^{t},\boldsymbol{\rm{P}}^{t};\xi^{t})$ and $\nabla_{\boldsymbol{\rm{P}}}r_{k}(\boldsymbol{\rm{A}}^{t},\boldsymbol{\rm{P}}^{t};\xi^{t})$ are the gradients of the instantaneous rate $r_{k}(\boldsymbol{\rm{A}},\boldsymbol{\rm{P}})$ with respect to $\boldsymbol{\rm{A}}$ and $\boldsymbol{\rm{P}}$ at the iteration $t$, respectively.
		
	{\bf Step 2:} In this step, by using the surrogate functions $\{\bar{f}_{i}^{t}(\boldsymbol{\rm{A}},\boldsymbol{\rm{P}}), \forall i\in 0\cup 1\cup\mathcal{O}\cup\mathcal{B}\}$, we express the convex approximation of problem (\ref{Problem_7}) at the iteration $t$ as
	\begin{equation}\label{Problem_7_1}
	\begin{aligned}
	&(\bar{\boldsymbol{\rm{A}}}^{t},\bar{\boldsymbol{\rm{P}}}^{t})={\arg\min}_{(\boldsymbol{\rm{A}},\boldsymbol{\rm{P}})\in\mathcal{X}}
	\quad\bar{f}_{0}^{t}(\boldsymbol{\rm{A}},\boldsymbol{\rm{P}})\\
	&\text{s.t.}\;\;~
	\bar{f}_{1}^{t}(\boldsymbol{\rm{A}},\boldsymbol{\rm{P}})\leq 0,
	&& \\
	& \quad\quad \bar{f}_{k}^{t}(\boldsymbol{\rm{A}},\boldsymbol{\rm{P}})\leq 0,
	&& \forall k \in \mathcal{O}, \\	
	& \quad\quad \bar{f}_{b}^{t}(\boldsymbol{\rm{A}},\boldsymbol{\rm{P}})\leq0,
	&& \forall b \in \mathcal{B}.
	\end{aligned}
	\end{equation}
	If problem (\ref{Problem_7_1}) is infeasible, the corresponding feasibility problem is
	\begin{equation}\label{Problem_8}
	\begin{aligned}
	&(\bar{\boldsymbol{\rm{A}}}^{t},\bar{\boldsymbol{\rm{P}}}^{t})={\arg\min}_{(\boldsymbol{\rm{A}},\boldsymbol{\rm{P}})\in\mathcal{X},\eta}\quad
	\eta\\
	& \text{s.t.}\;\;~
	\bar{f}_{1}^{t}(\boldsymbol{\rm{A}},\boldsymbol{\rm{P}})\leq \eta,
	&& \\
	&\quad\quad \bar{f}_{k}^{t}(\boldsymbol{\rm{A}},\boldsymbol{\rm{P}})\leq \eta,
	&& \forall k \in \mathcal{O}, \\	
	&\quad\quad \bar{f}_{b}^{t}(\boldsymbol{\rm{A}},\boldsymbol{\rm{P}})\leq \eta,
	&& \forall b\in \mathcal{B}.
	\end{aligned}
	\end{equation}
	Problems (\ref{Problem_7_1}) and (\ref{Problem_8}) are convex. Hence, they can be solved optimally by using existing convex optimization solvers, such as CVX \cite{CVX} --\cite{Boyd2004}.
	
	{\bf Step 3:} Given the optimal solution $\bar{\boldsymbol{\rm{A}}}^{t}$ and $\bar{\boldsymbol{\rm{P}}}^{t}$ in one of the above two cases, $\boldsymbol{\rm{A}}$ and $\boldsymbol{\rm{P}}$ are updated according to
	\begin{equation}
	\begin{aligned} \label{update_new_1}
	\boldsymbol{\rm{A}}^{t+1} = (1-\beta^t)\boldsymbol{\rm{A}}^t + \beta^t\bar{\boldsymbol{\rm{A}}}^t,
	\end{aligned}
	\end{equation}
	and
	\begin{equation}
	\begin{aligned}\label{update_new_2}
	\boldsymbol{\rm{P}}^{t+1} = (1-\beta^t)\boldsymbol{\rm{P}}^t + \beta^t\bar{\boldsymbol{\rm{P}}}^t,
	\end{aligned}
	\end{equation}
	respectively (as shown in Fig. \ref{BD}), where the step size $\beta^t\in(0, 1]$ is a sequence that needs to be chosen accurately.
	\subsection{Convergence and Computational Complexity Analysis}
	In this subsection, we prove the convergence of the proposed Algorithm \ref{SHA} to a stationary point and also calculate its computational complexity. 
	
	The convergence conditions of the CSSCA method are provided in \cite{Liu2019} and \cite{Liu2018}. In this subsection, we show that the sufficient conditions for the convergence of the proposed algorithm are satisfied. To achieve this, we confirm that the adopted surrogate functions $\bar{f_{i}^{t}}(\boldsymbol{\rm{A}},\boldsymbol{\rm{P}})$ hold properties that are essential for convergence as explained in the following. First, the following assumptions have to be made on the step sizes $\rho^{t}$ and $\beta^{t}$.
	\begin{assumption} \label{Ass_1}
		We define the following assumptions on the sequence of step sizes $\rho^{t}$ and $\beta^{t}$:
		\begin{enumerate}
			\item $\rho^{t}\rightarrow0$, $\sum_{t} \rho^{t}=\infty$, $\sum_{t}{(\rho^{t})}^{2}<\infty$, $\lim_{t\rightarrow\infty}\rho^{t}t^{-1/2}<\infty$,
			\item $\beta^{t}\rightarrow0$, $\sum_{t} \beta^{t}=\infty$, $\sum_{t}{(\beta^{t})}^{2}<\infty$,
			\item $\lim_{t\rightarrow\infty}{\beta^{t}}/{\rho^{t}}=0$.
		\end{enumerate}
	\end{assumption}
	Now, based on Assumption \ref{Ass_1}, we can observe that the adopted surrogate functions, i.e., $\bar{f}_{i}^{t}(\boldsymbol{\rm{A}},\boldsymbol{\rm{P}})$, have the following properties described in Lemma \ref{lem_3}.
	\begin{lem}\label{lem_3}
		For all $t = \{0,1,...\}$, the properties of the surrogate function $\bar{f_{i}}^{t}(\boldsymbol{\rm{A}},\boldsymbol{\rm{P}})$ are as follows
		\begin{enumerate}
			\item {
				$\bar{f_{i}}^{t}(\boldsymbol{\rm{A}},\boldsymbol{\rm{P}})$ is uniformly strongly convex in $\boldsymbol{\rm{A}}$ and $\boldsymbol{\rm{P}}$.
			}
			\item {
				$\bar{f_{i}}^{t}(\boldsymbol{\rm{A}},\boldsymbol{\rm{P}})$ is a Lipschitz continuous function with respect to $\boldsymbol{\rm{A}}$ and $\boldsymbol{\rm{P}}$. Furthermore, $\lim\sup_{t_{1},t_{2}\rightarrow\infty} \bar{f_{i}}^{t_{1}}(\boldsymbol{\rm{A}},\boldsymbol{\rm{P}}) - \bar{f_{i}}^{t_{2}}(\boldsymbol{\rm{A}},\boldsymbol{\rm{P}})\leq K (||\boldsymbol{\rm{A}}^{t_{1}} - \boldsymbol{\rm{A}}^{t_{2}}||+||\boldsymbol{\rm{P}}^{t_{1}} - \boldsymbol{\rm{P}}^{t_{2}}||), \forall (\boldsymbol{\rm{A}},\boldsymbol{\rm{P}})\in\mathcal{X}$ for some constant $K > 0$.
			}
			\item {
				For any $(\boldsymbol{\rm{A}},\boldsymbol{\rm{P}})\in\mathcal{X}$, $\bar{f_{i}}^{t}(\boldsymbol{\rm{A}},\boldsymbol{\rm{P}})$, its derivative, and its second-order derivative are uniformly bounded.
			}
			\item {
				$\lim_{t\rightarrow\infty}|\bar{f}_{i}^{t}(\boldsymbol{\rm{A}}^{t},\boldsymbol{\rm{P}}^{t}) - f_{i}(\boldsymbol{\rm{A}}^{t},\boldsymbol{\rm{P}}^{t})| = 0$ and $\lim_{t\rightarrow\infty}\allowbreak||\nabla\bar{f_{i}}^{t}(\boldsymbol{\rm{A}}^{t},\boldsymbol{\rm{P}}^{t}) - \nabla f_{i}(\boldsymbol{\rm{A}}^{t},\boldsymbol{\rm{P}}^{t})|| = 0$.
			}
		\end{enumerate}
	\end{lem}
	\begin{proof}
	The surrogate functions $\bar{f_{i}}^{t}(\boldsymbol{\rm{A}},\boldsymbol{\rm{P}})$ are a special case of the structured surrogate function with a zero convex component in \cite{Liu2019}. As a result, under Assumption \ref{Ass_1}, all properties mentioned in this lemma are met for all the adopted surrogate functions.
	\end{proof}
	\begin{thm} \label{thm_3}
		(Convergence of Algorithm \ref{SHA}): Suppose that Assumption \ref{Ass_1} and Lemma \ref{lem_3} are satisfied, and the initial point $\{\boldsymbol{\rm{A}}^{0},\boldsymbol{\rm{P}}^{0}\}$ is a feasible point. Let $\{\boldsymbol{\rm{A}}^{t},\boldsymbol{\rm{P}}^{t}\}_{t=1}^{\infty}$ denote the iterates generated by Algorithm \ref{SHA}. Then, every limiting point $\{\boldsymbol{\rm{A}}^{*},\boldsymbol{\rm{P}}^{*}\}$ of $\{\boldsymbol{\rm{A}}^{t},\boldsymbol{\rm{P}}^{t}\}_{t=1}^{\infty}$ satisfying the Slater condition is a stationary point of problem (\ref{Problem_5}) almost surely.
	\end{thm}
	\begin{proof}
	It has been shown in \cite{Liu2019} --\cite{Liu2018} that under Lemma \ref{lem_3} and Assumption \ref{Ass_1}, structured surrogate functions \cite{Yang2016} converge to a stationary point. Due to the fact that our adopted surrogate functions are structured surrogate functions and Lemma \ref{lem_3} and Assumption \ref{Ass_1} are satisfied, Algorithm \ref{SHA} converges to a stationary point. Thus, this theorem is proven.
	\end{proof}
	We analyze the computational complexity of the proposed Algorithm \ref{SHA} for each iteration in the following. As mentioned above, we solve the optimization problems (\ref{Problem_7_1}) and (\ref{Problem_8}) by the CVX toolbox; CVX uses the interior point method \cite{CVX} --\cite{Boyd2004}. The optimization problems (\ref{Problem_7_1}) and (\ref{Problem_8}) consist of $C=2B+S+SB+2L+LB+1$ convex constraints, and thus, the number of iterations required for the interior point method is obtained by $\dfrac{\log\left((C)/(i^{0}\Lambda)\right)}{\log(\varsigma)}$ \cite{Nemirovski}, where $i^{0}$ is the initial point, $0<\Lambda<\!\!<1$ is the stopping criterion, and $\varsigma$ is used to update the accuracy of the interior point method. With a polynomial presentation, the number of required iterations grows as $\sqrt{C}$ (asymptotically $\approx \sqrt{LB}$) \cite{CVX}, \cite{Nemirovski}. Moreover, for each iteration, additional computational complexity is incurred to update the surrogate functions employing the CSSCA method in step 4, which is $O((S+B)LB)$ \cite{Liu2018}. Therefore, the computational complexity of the proposed Algorithm \ref{SHA} at each iteration is $O((S+B)(LB)^{1.5})$, which is polynomial time complexity.
	\section{Numerical Results}\label{Res}
	In this section, we evaluate the convergence and performance of the proposed algorithm with extensive numerical results. To this end, we consider a  cellular network with $S$ seller MNOs and $B$ buyer MNOs. The BSs and users of MNOs are spatially distributed according to independent PPPs inside a circular area with a 500 m radius. Moreover, we assume that all MNOs have the same intensity of BSs and users per unit area (i.e., $\lambda_{s}=\lambda_{b}$ and $\mu_{s}=\mu_{b}, \forall b\in\mathcal{B}, \forall s\in\mathcal{S}$). The transmit power of each buyer BS on the licensed sub-bands is obtained by (\ref{Power}). The other simulation parameters are given in Table \ref{tab:parameters}. Note that in all scenarios, the numerical results are obtained by averaging over 500 instances of random parameters for the stochastic configurations.
	\begin{table}[t]
		\caption{Simulation Parameters}
		\label{tab:parameters}
		\centering
		\footnotesize
		\renewcommand{\arraystretch}{1.0}
		\begin{tabular}{|l|l|}
			\cline{1-2}
			\textbf{Parameter}&\textbf{Value} \\
			\hline
			\hline
			radius of circular area ($r$)& 500 m \cite{Sanguanpuak2017Conf}, \cite{Sanguanpuak2017}\\
			\hline
			intensity of BSs per MNO ($\lambda_{k}$)&8 $\text{BS/m}^{2}$ \cite{Sanguanpuak2017Conf}, \cite{Sanguanpuak2017}\\
			\hline
			intensity of users per MNO ($\mu_{k}$)&16 $\text{UEs/m}^{2}$\\
			\hline
			\begin{tabular}{@{}l@{}}maximum transmit power of\\ each sellers' BS ($P^{\text{max}}$)\end{tabular}& 10 dBm \cite{Sanguanpuak2017Conf}\\
			\hline
			maximum interference threshold ($\zeta_{s}$)& -110 dBm \cite{Gupta2016}\\
			\hline
			path-loss exponent ($\alpha$)&4 \cite{Sanguanpuak2017Conf}, \cite{Sanguanpuak2017}\\
			\hline
			noise power ($\sigma^{2}$)& -150 dBm \cite{Sanguanpuak2017Conf}\\
			\hline
			price per 1bps/Hz ($\delta$)& 2 $\$$ \cite{Cano2016}\\
			\hline
			investment time ($D$)& 120 months \cite{Cano2016}\\
			\hline
			price coefficient of the first seller MNO ($\varphi_{1,b}$)&1800 $\$$\\
			\hline
			minimum data rate requirement (${r^{\text{th}}}$)&1 bps/Hz\\
			\hline
			\begin{tabular}{@{}l@{}}license price per sub-band paid by seller\\ MNO $s$ to regulators ($\varphi_{s}$)\end{tabular} & 2000 $\$$ \cite{Cano2016}\\
			\hline
			$v_s$ for all $s\in\mathcal{S}$& $1/S$\\
			\hline
			$w_b$ for all $b\in\mathcal{B}$& $1/B$\\
			\hline
			price coefficient of the second seller MNO ($\varphi_{2,b}$)&1200 $\$$\\
			\hline number of samples ($T$)&500\\
			\cline{1-2}
		\end{tabular}
	\end{table}
	\begin{figure}[!t]	
		\centering
		\includegraphics[width=9.5 cm,height=5.5 cm]{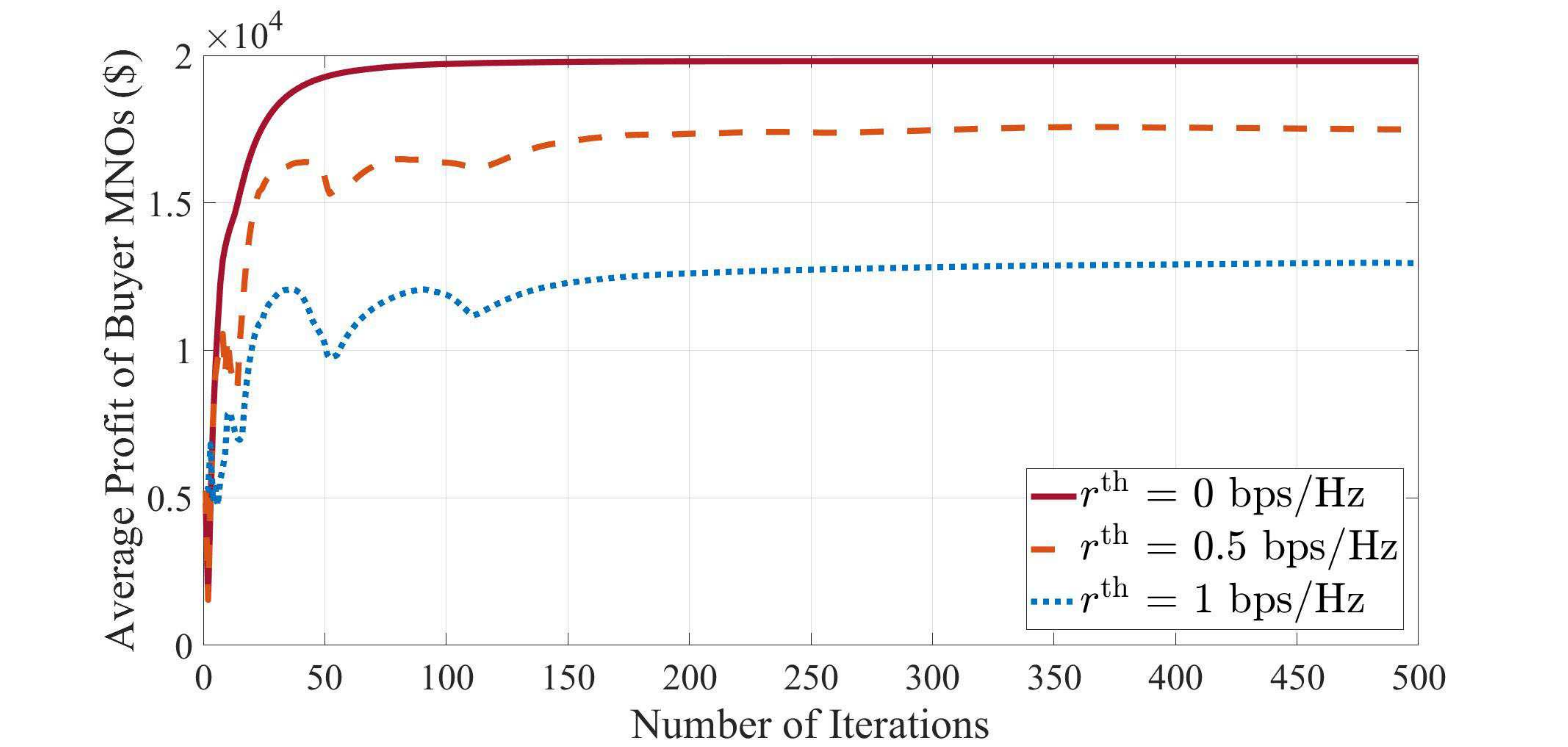}
		\caption{Convergence of the proposed algorithm for different values of $r^{\text{th}}$.\label{demo1}}
	\end{figure}
	\begin{figure}[!t]
		\centering
		{\includegraphics[width=9.5 cm,height=5.5 cm]{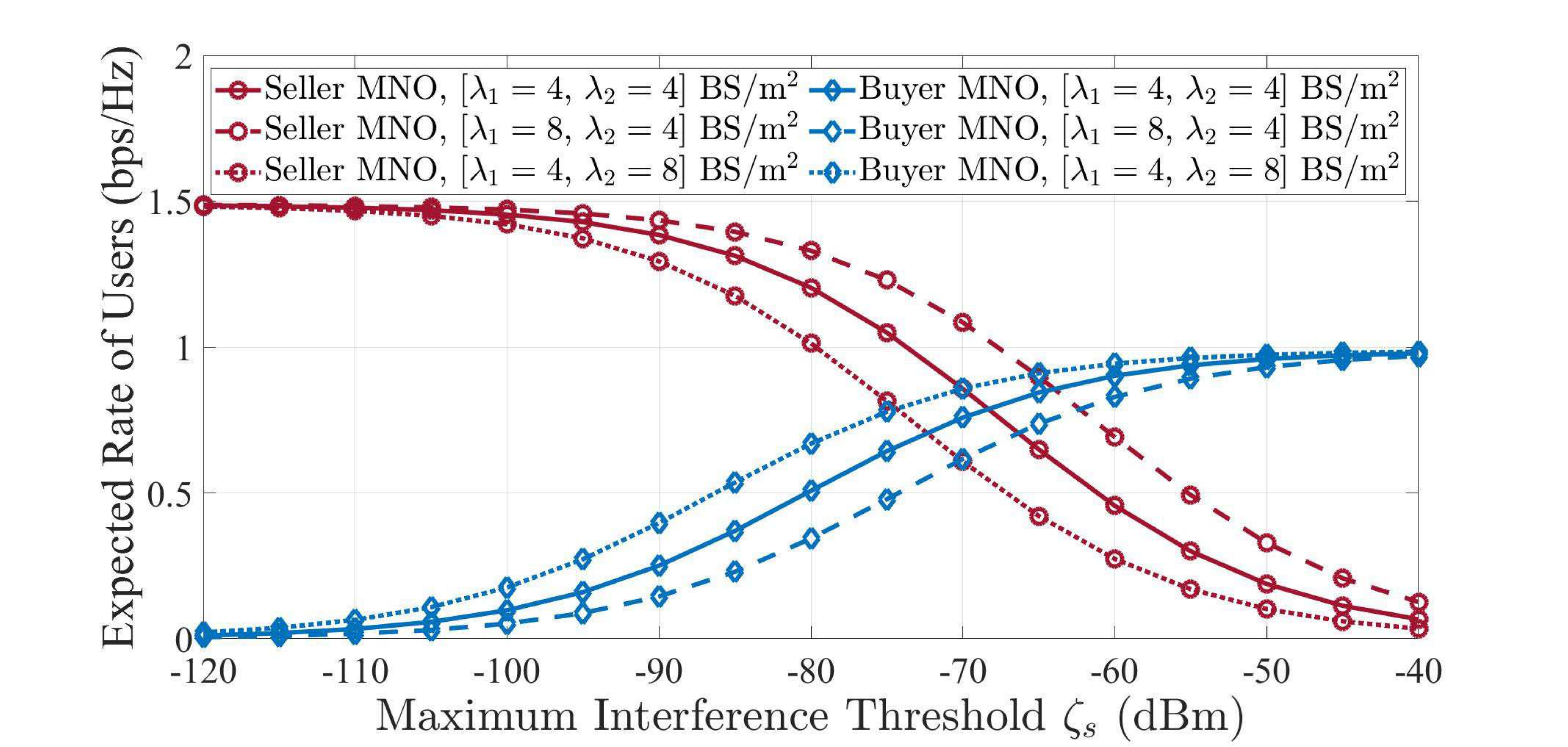}
			\caption{Expected rate of the seller and buyer users versus the maximum interference threshold for different values of BS intensity.\label{demo2}}}
	\end{figure}
	\subsection{Convergence of the Proposed Algorithm}
	Fig. \ref{demo1} shows the convergence of the proposed algorithm in terms of the number of iterations (or samples) for different values of minimum data rate requirements ($r^{\text{th}}$). For generating this figure, we set the number of seller MNOs to 2 ($S=2$), the number of licensed sub-bands per seller MNO to 1 ($L_{s} = 1$), and the number of buyer MNOs to 1 ($B=1$). It can be observed that the proposed algorithm converges to a stationary point. From Fig. \ref{demo1}, we also see that increasing the minimum data rate requirement of users brings about the average profit of the buyer MNOs to decrease. In fact, the average profit of the buyer MNOs has the highest value for the case of $r^{\text{th}} = 0$ bps/Hz. As a result of the increasing $r^{\text{th}}$, the buyer MNOs must buy more licensed sub-bands from seller MNOs to guarantee $r^{\text{th}}$ for their users. Therefore, buyer MNOs have to pay more costs, which results in a decreased profit of the buyer MNOs based on (\ref{u_b}).
		\begin{figure}[!t]
		\centering
		{\includegraphics[width=9.5 cm,height=5.5cm]{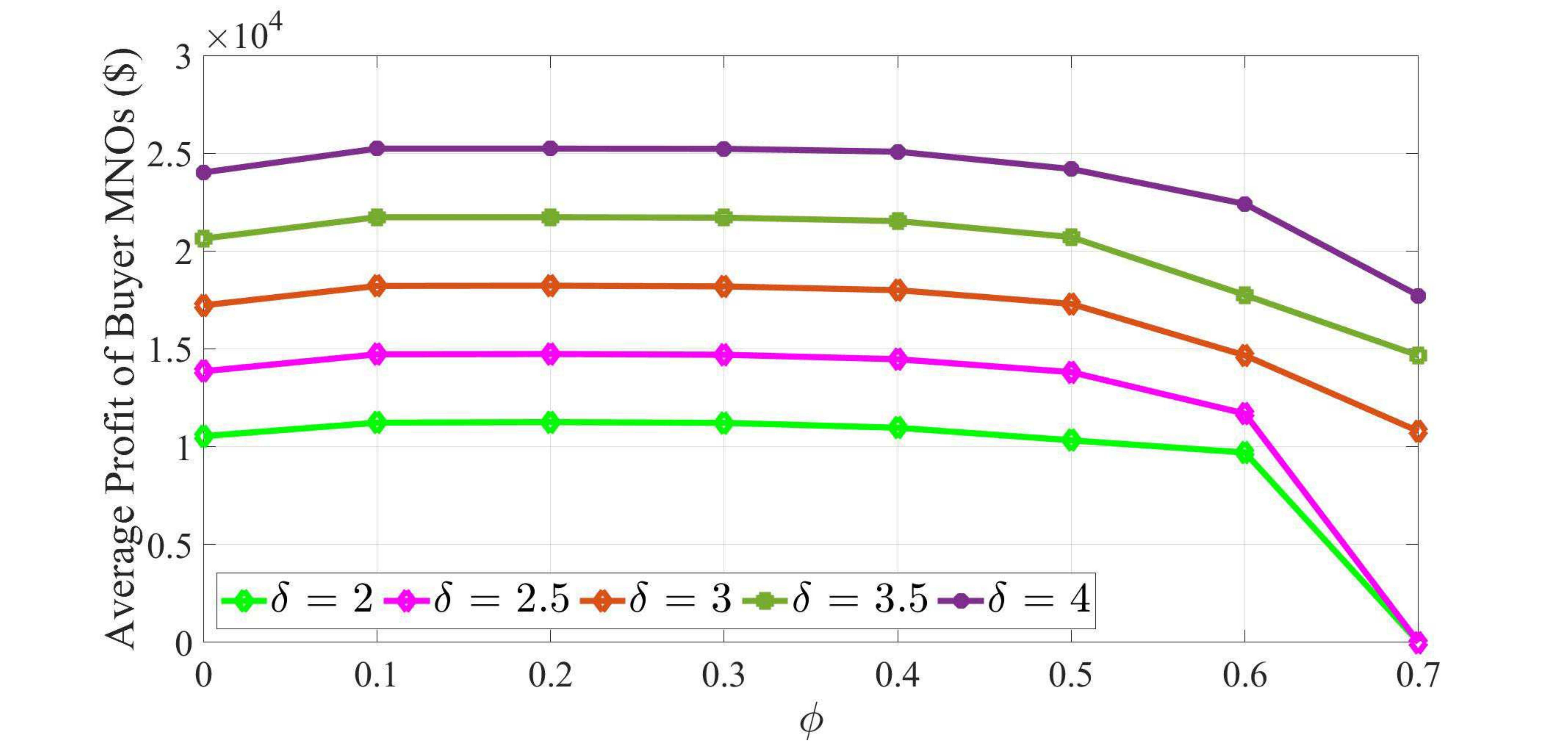}
			\caption{Average profit of the buyer MNOs versus $\phi$ for different values of $\delta$.\label{demodelta}}}
	\end{figure}
	\begin{figure}[!t]
		\centering
		{\includegraphics[width=9.5 cm,height=5.5 cm]{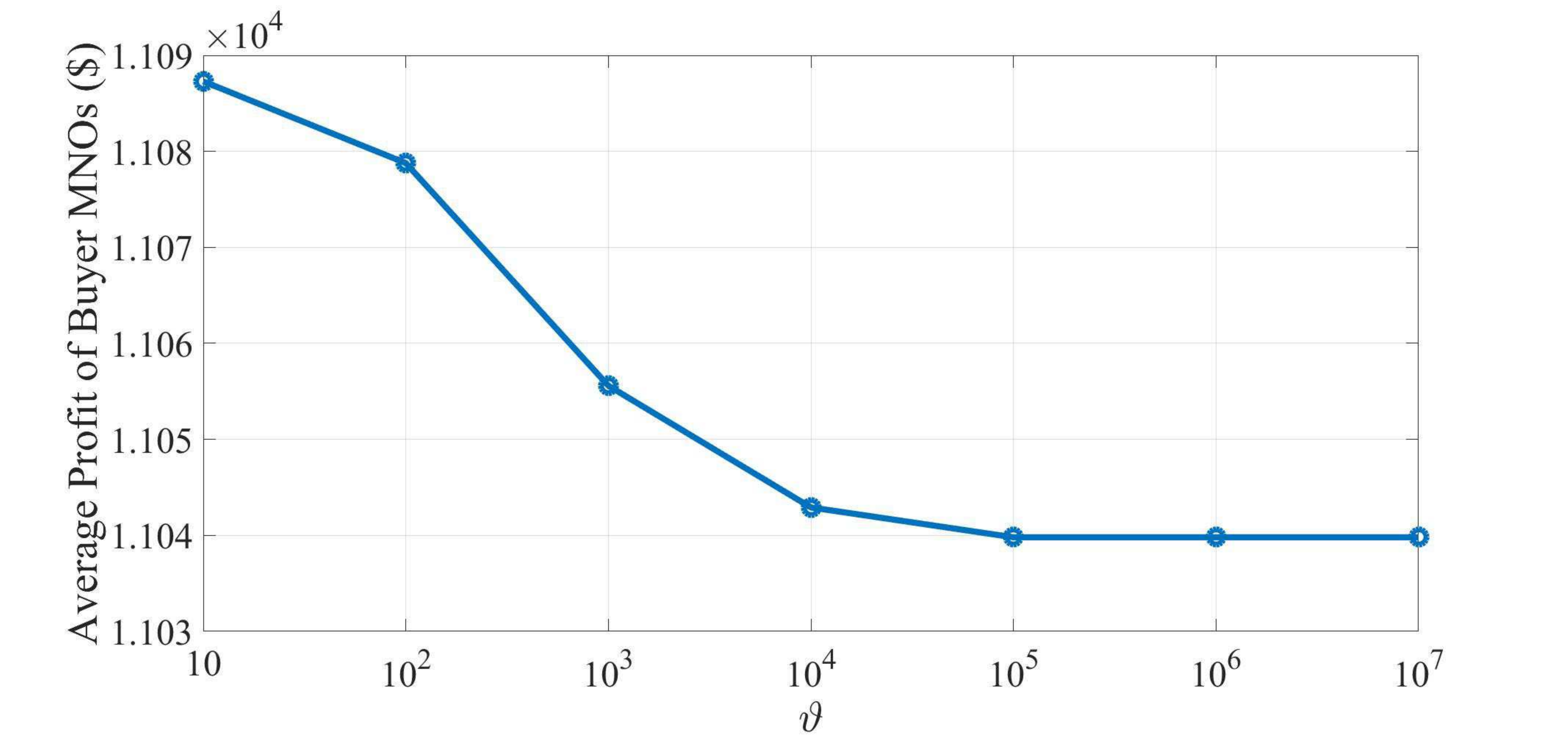}
			\caption{Average profit of the buyer MNOs versus $\vartheta$.\label{demov}}}
	\end{figure}
	\subsection{Effect of Different Parameters on the Performance of the Proposed Algorithm}
	In Fig. \ref{demo2}, the expected rate of both the seller and buyer users is demonstrated versus the maximum interference threshold $\zeta_{s}$ for different values of BS intensity. For generating this figure, we consider a single seller and buyer MNO, where the seller MNO has one licensed sub-band, as in \cite{Gupta2016}. From Fig. \ref{demo2}, we can observe that with an increasing value of the buyer MNO's BS intensity ($\lambda_{2}$), the per-user expected rate of the buyer MNO improves. Moreover, because of the increasing interference caused by the buyer MNO's BSs, the seller MNO's per-user expected rate decreases. On the other hand, with an increasing value of the seller's BS intensity ($\lambda_{1}$), we can observe that the per-user expected rate of the seller MNO improves. From Fig. \ref{demo2}, it can be seen that increasing the maximum interference threshold ($\zeta_{s}$) leads to a decrease in the per-user expected rate of the seller MNO and an increase in the buyer MNO's per-user expected rate. The reason is that when the maximum interference threshold increases, the transmit power of the buyer MNO's BSs based on (\ref{Power}) can increase to get a higher per-user expected rate. Because the transmit power of the seller's BSs is constant, when the maximum interference threshold ($\zeta_{s}$) increases, the seller MNO's per-user expected rate decreases according to Lemma \ref{thm_2}.
		
	Fig. \ref{demodelta} illustrates the average profit of the buyer MNOs versus the trade-off parameter $\phi$ in (\ref{phi}) for different values of $\delta$. To generate this figure, we set the number of seller and buyer MNOs to 2 ($S = 2$ and $B = 2$), and each seller MNO has only one licensed sub-band ($L_{s} = 1$). As can be seen in Fig. \ref{demodelta}, with an increasing $\delta$, the average profit of the buyer MNOs increases. This is due to the fact that buyer MNOs' users must pay more cost to their MNOs based on (\ref{rev_m}). Furthermore, when $\delta$ is fixed, the average profit of the buyer MNOs first increases with $\phi$ until reaching the peak and then decreases with the growth of $\phi$. When $\phi$ is small, the required transmit power of seller MNOs' BSs to meet constraints C0 and C1 is negligible. Hence, the buyer MNOs can tolerate less interoperator interference from the seller MNOs, and thus, the average profit of the buyer MNOs increases. After that, the growth of $\phi$ leads to an increasing transmit power of the seller MNOs' BSs, which is not negligible. Consequently, the interoperator interference imposed on buyer users increases, and the average profit of the buyer MNOs turns to decrease.
	\begin{figure}[!t]
		\centering
		{\includegraphics[width=9.5 cm,height=5.5 cm]{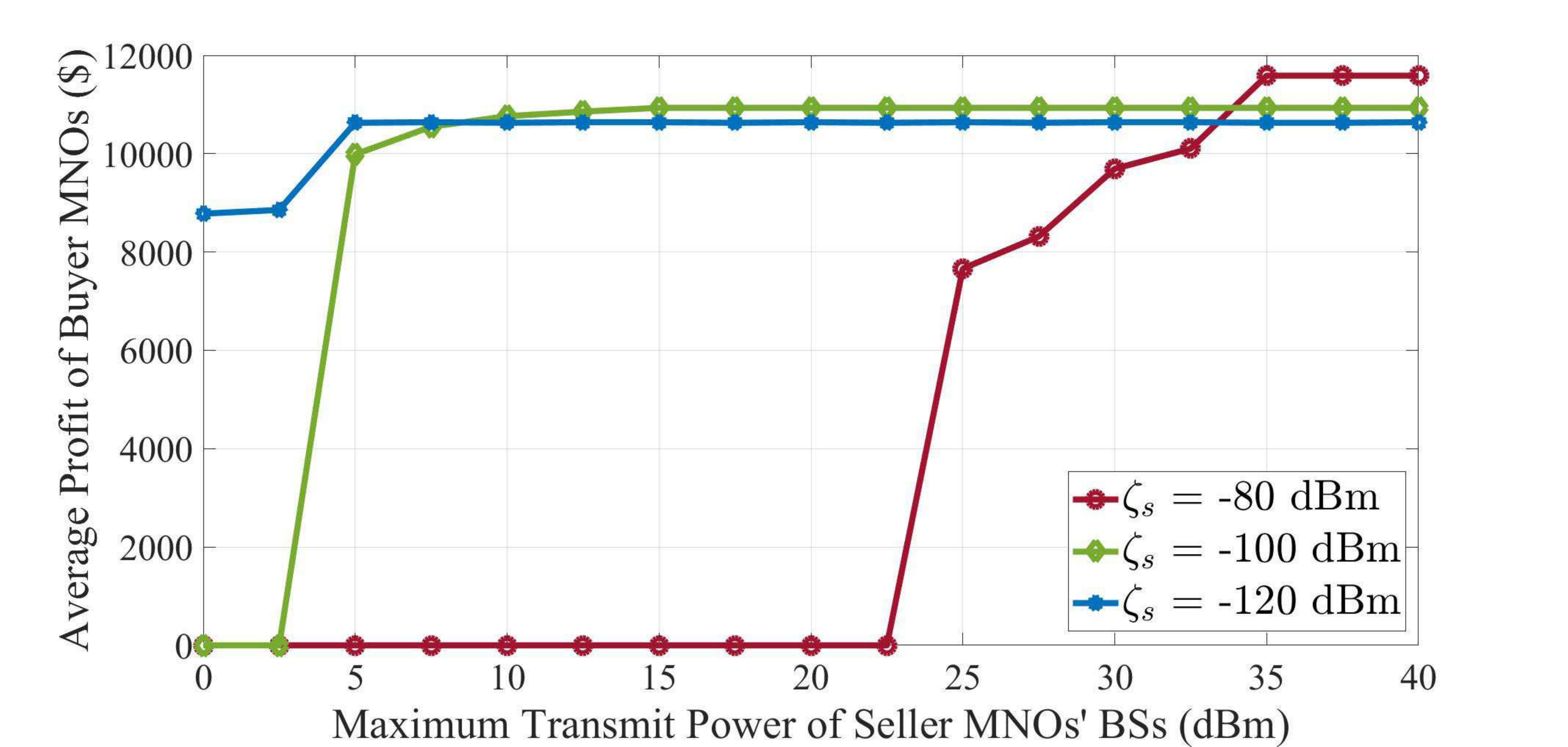}
			\caption{Average profit of the buyer MNOs versus $P^{\text{max}}$ for different values of $\zeta_{s}$ for all $s\in\mathcal{S}$.\label{demoP}}}
	\end{figure}
	\begin{figure}[!t]
		\centering
		\includegraphics[width=9.5 cm,height=5.5 cm]{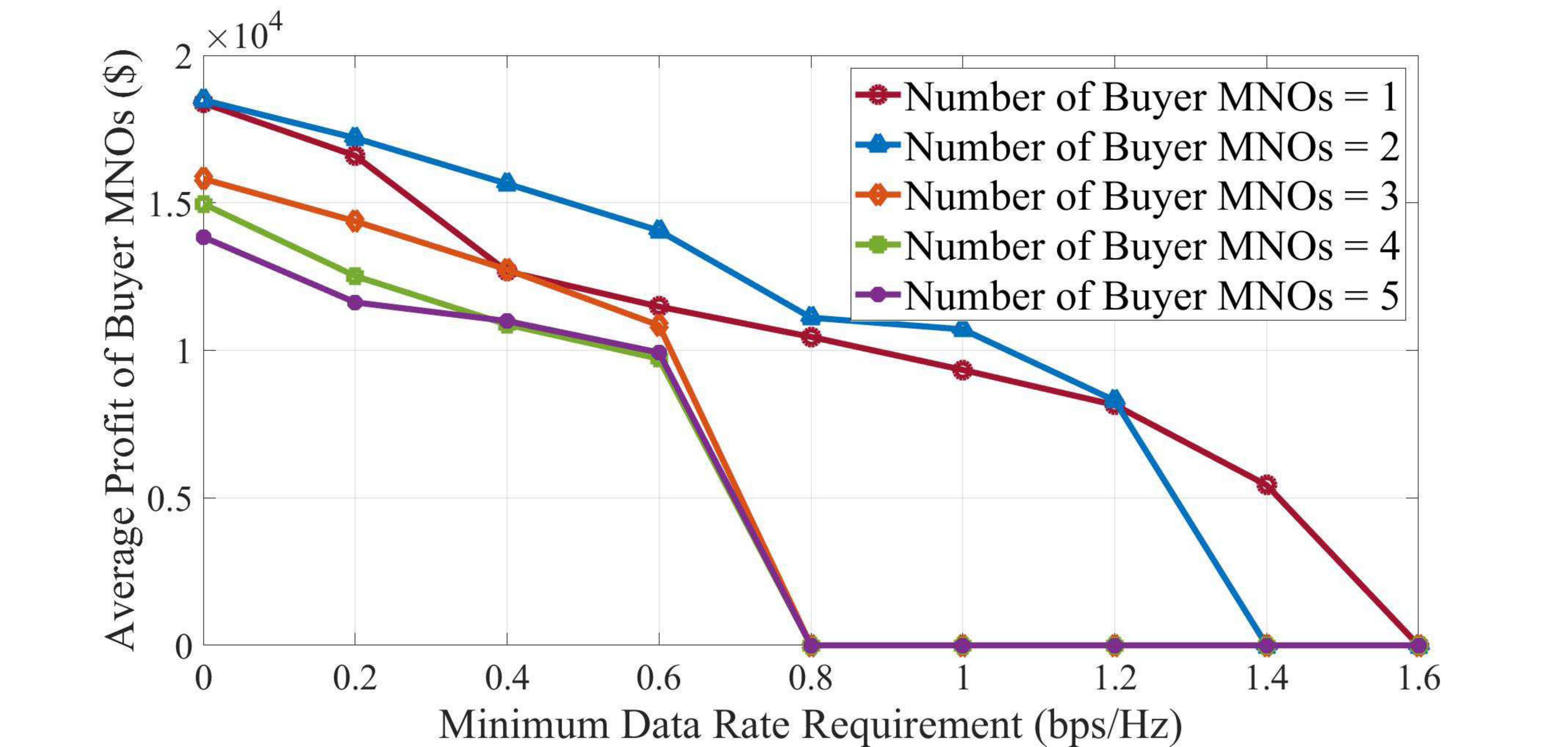}
		\caption{Average profit of the buyer MNOs versus different values of $r^{\text{th}}$ for different numbers of buyer MNOs ($B$). \label{demo5}}
	\end{figure}

	The average profit of the buyer MNOs versus different values of $\vartheta$ is plotted in Fig. \ref{demov}, where the number of seller MNOs ($S$) and the number of buyer MNOs ($B$) are set to 2 ($S = 2$ and $B = 2$), and each seller MNO has only one licensed sub-band ($L_{s} = 1$). The parameter $\vartheta$ is a penalty factor to penalize the average profit of the buyer MNOs in problem (\ref{Problem_5}) when the values of sub-band sharing variables, i.e., $a_{l_{s},b}$, are neither 0 nor 1. In Fig. \ref{demov}, we can observe that when $\vartheta$ is small, the cost of violation of $a_{l_{s},b}-a_{l_{s},b}^2=0$ becomes low. Thus, the binary nature of some sub-band sharing variables can be clearly violated, i.e., $a_{l_{s},b}-a_{l_{s},b}^2=0$ does not hold. In this case, a proportion of sub-band $l_{s}$ is assigned to one buyer MNO, while the remaining proportion can be assigned to others, which leads to an increasing profit of the buyer MNOs. As  $\vartheta$ increases, the cost of violation of $a_{l_{s},b}-a_{l_{s},b}^2=0$ grows, which forces each sub-band sharing variable to get closer to 0 or 1. Consequently, the average profit of the buyer MNOs decreases, as the number of sub-bands that an MNO can lease to serve its users would be restricted. However, after reaching a specific value, here $10^5$, the average profit of the buyer MNOs converges to its final value, remaining stable. The reason is that when $\vartheta$ is sufficiently large, the penalty term converges to zero.
	
	Fig. \ref{demoP} demonstrates the average profit of the buyer MNOs versus the maximum transmit power of seller MNOs' BSs ($P^{\text{max}}$) for different values of maximum interference threshold ($\zeta_{s}$). For generating this figure, the number of both seller ($S$) and buyer ($B$) MNOs is set to 2 ($S = 2$ and $B = 2$), and each seller MNO has only one licensed sub-band ($L_{s} = 1$). As observed in Fig. \ref{demoP}, for some small values of $P^{\text{max}}$, MNOs are unable to guarantee the minimum data rate requirement of their users, which leads to an infeasible system. \footnote{Average profit of zero means that the proposed algorithm is infeasible.} Nonetheless, the rate of increase in the average profit of the buyer MNOs is increased as the maximum transmit power becomes larger until the average profit of the buyer MNOs reaches a constant value in the high transmit power regime. In fact, buyer MNOs attempt to maximize their average profit function by leasing licensed sub-channels from seller MNOs and by providing the minimum data rate requirement for their users, resulting in an increasing inter-operator interference imposed on each user of the seller MNOs. Therefore, increasing the maximum transmission power allows the seller MNOs to share their sub-channels and adjust the transmission power of their BSs simultaneously so that the minimum data rate requirement of their users is met and additional revenue is achieved. When the maximum average profit of the buyer MNOs is obtained, a further increase in $P^{\text{max}}$ has no advantage for MNOs. Moreover, in Fig. \ref{demoP} we see that for a constant transmit power of seller MNOs' BSs, the average profit of the buyer MNOs increases with an increasing $\zeta_{s}$. The reason is that as $\zeta_{s}$ increases, the transmit power of the buyer MNOs' BSs based on (\ref{Power}) increases, leading to a higher per-user expected rate for the buyer MNOs. Hence, the average profit of the buyer MNOs increases.
		\begin{figure}[!t]
		\centering
		\includegraphics[width=9.5 cm,height=5.5 cm]{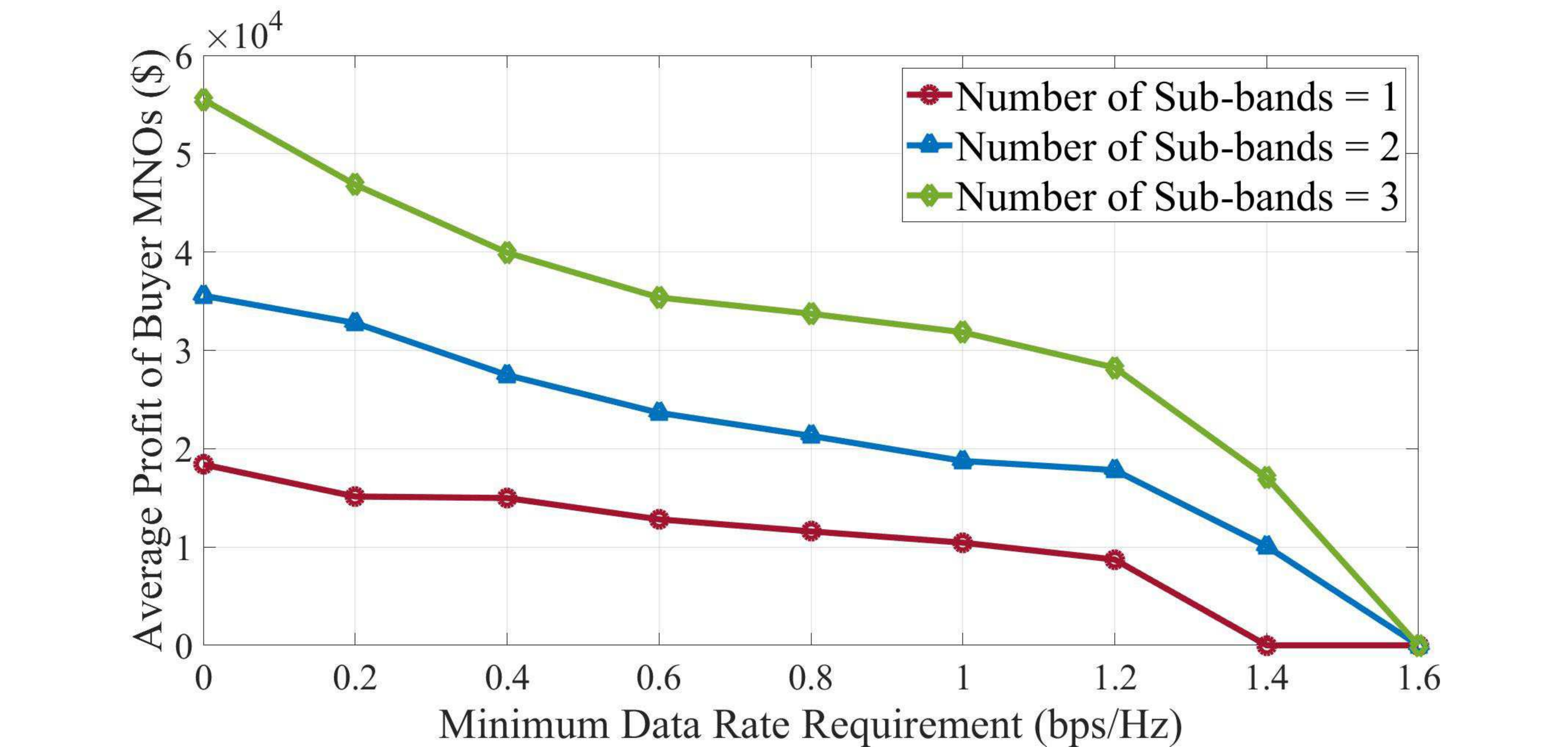}
		\caption{Average profit of the buyer MNOs versus different values of $r^{\text{th}}$ for different numbers of sub-bands per seller MNO ($L_{s}, s\in\mathcal{S}$).\label{demo6}}

	\end{figure}
	\begin{figure}[!t]
		\centering
		\includegraphics[width=9.5 cm,height=5.5 cm]{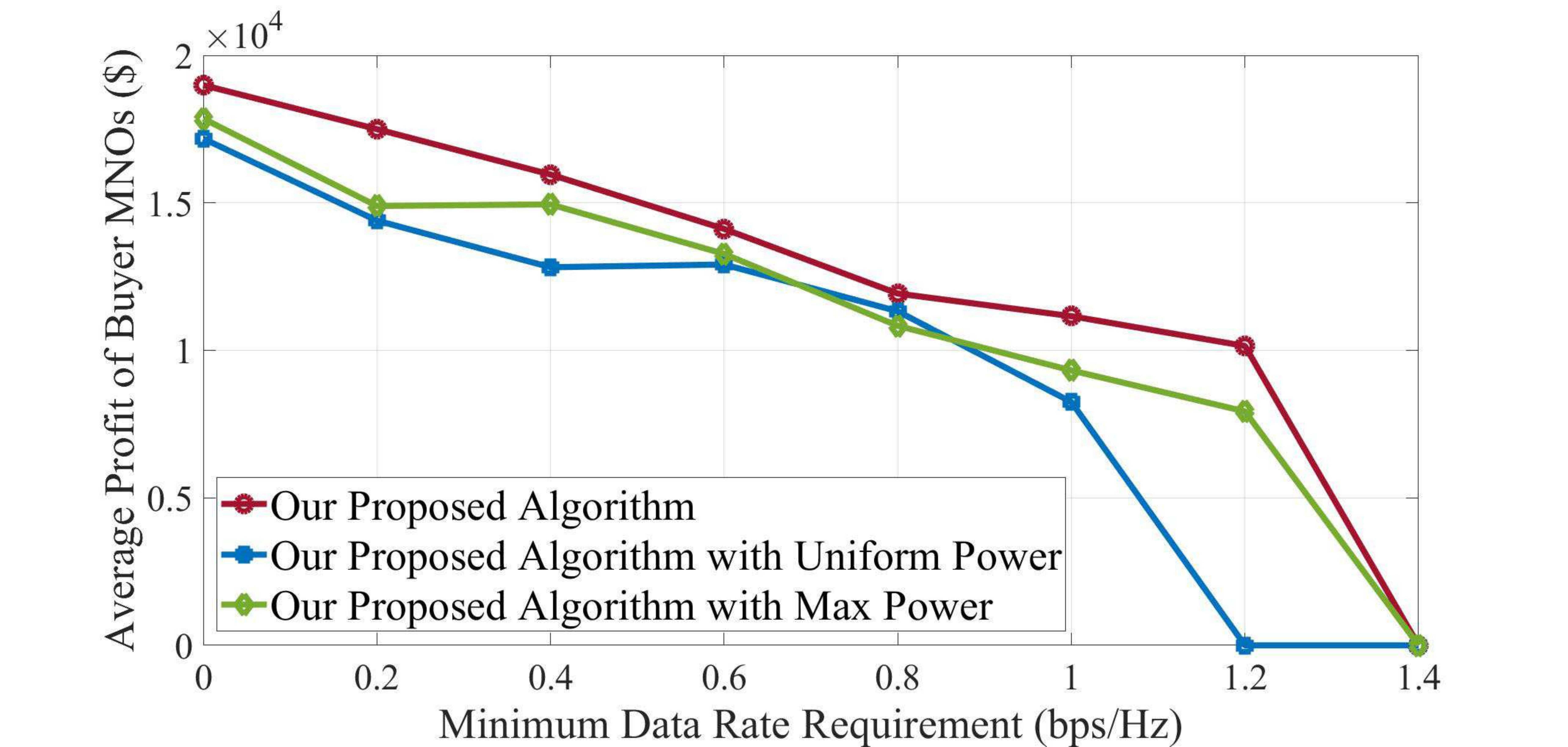}
		\caption{Average profit of all MNOs versus different values of $r^{\text{th}}$ for different power allocation schemes.\label{demo7}}
	\end{figure}

	The average profit of the buyer MNOs versus different values of minimum data rate requirements ($r^{\text{th}}$) for different numbers of buyer MNOs ($B$) is illustrated in Fig. \ref{demo5}. To generate this figure, we set the number of seller MNOs to 2 ($S=2$), and each seller MNO has only one licensed sub-band ($L_{s}=1$). From Fig. \ref{demo5}, we observe that with an increasing $r^{\text{th}}$, the profit of the buyer MNOs decreases. Because of the increasing $r^{\text{th}}$, the buyer MNOs must buy more licensed sub-bands from sellers to guarantee $r^{\text{th}}$ for their users. Therefore, the buyer MNOs have to pay more costs, which leads to a decreased profit of the buyer MNOs based on (\ref{u_b}). However, for some values of $r^{\text{th}}$, the buyer MNOs are unable to guarantee $r^{\text{th}}$ for their users by shared sub-bands, leading to an infeasible system. Furthermore, it can be seen that increasing the number of buyer MNOs leads to an increased average profit of the buyer MNOs. As a result of increasing the number of buyer MNOs, the number of users simultaneously served by MNOs increases. Therefore, according to (\ref{rev_m}), the average profit of the buyer MNOs increases. However, after some points of $B$ (e.g., $B = 3$), the average profit of the buyer MNOs decreases because the interference on buyer users is intensified. Another interesting observation is that when increasing the number of buyer MNOs, the system rapidly goes to an infeasible state as the users' interference is intensified.
	
	Fig. \ref{demo6} shows the average profit of the buyer MNOs versus different values of minimum data rate requirements ($r^{\text{th}}$) for different numbers of sub-bands per seller MNO ($L_{s}$). To generate this figure, the number of both seller ($S$) and buyer MNOs ($B$) is set to 2. Similar to Fig. \ref{demo5}, we can see that the average profit decreases when the minimum data rate requirement increases. Further, as shown in Fig. \ref{demo6}, the average profit of the buyer MNOs increases with an increasing number of leased sub-bands per seller MNO. As a result of increasing the number of leased sub-bands, the per-user expected rate increases, leading to an increased average profit. It is worth noticing that increasing the number of leased sub-bands increases the probability of feasibility.
	\subsection{Effect of Different Power Allocation Schemes on the Performance of the Proposed Algorithm}
	In this subsection, we evaluate the performance of the proposed algorithm in three cases in Fig. \ref{demo7}. In the first case, we adopt the transmit power of the buyer MNOs' BSs based on the power control strategy introduced in subsection \ref{Power_Strategy}. In the second case, we set the transmit power of the buyer MNOs' BSs based on the uniform power allocation in $[0,10{\text{ dBm}}]$. In the third one, we assume that the transmit power of the buyer MNOs' BSs is equal to the maximum transmit power of the seller MNOs' BSs (i.e., $p_{b,l_{s}} = P^{\text{max}} = 10$ dBm). For generating Fig. \ref{demo7}, we set the number of seller MNOs and buyer MNOs to 2 (i.e., $S = 2$ and $B = 2$), and each seller MNO has only one licensed sub-band ($L_{s}=1$). From Fig. \ref{demo7}, we can observe that the first case outperforms the other cases. The reason is that in the first case, the adopted power control strategy justifies the transmit power of the buyer MNOs' BSs. Therefore, the interference that each buyer MNO imposes on seller MNOs and other buyer MNOs is controlled, which leads to an increase in the average profit of all MNOs.
	\begin{figure}[!t]
		\begin{subfigure}{.95\columnwidth}
			\centering
			\includegraphics[width=9.5 cm,height=5.5 cm]{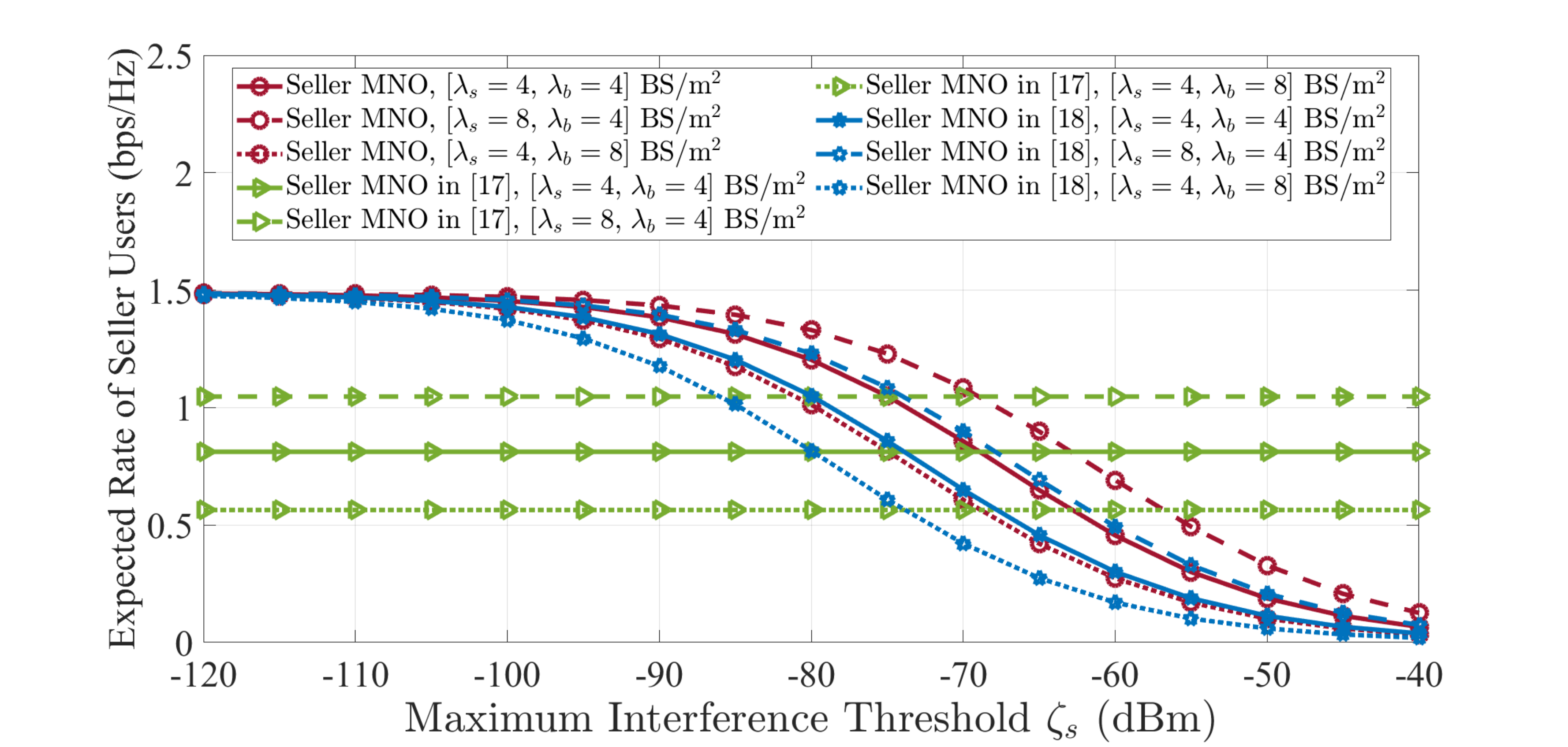}
			\caption{}
			\label{fig:sub-first}
		\end{subfigure}
	\begin{subfigure}{.95\columnwidth}
		\centering
		\includegraphics[width=9.5 cm,height=5.5 cm]{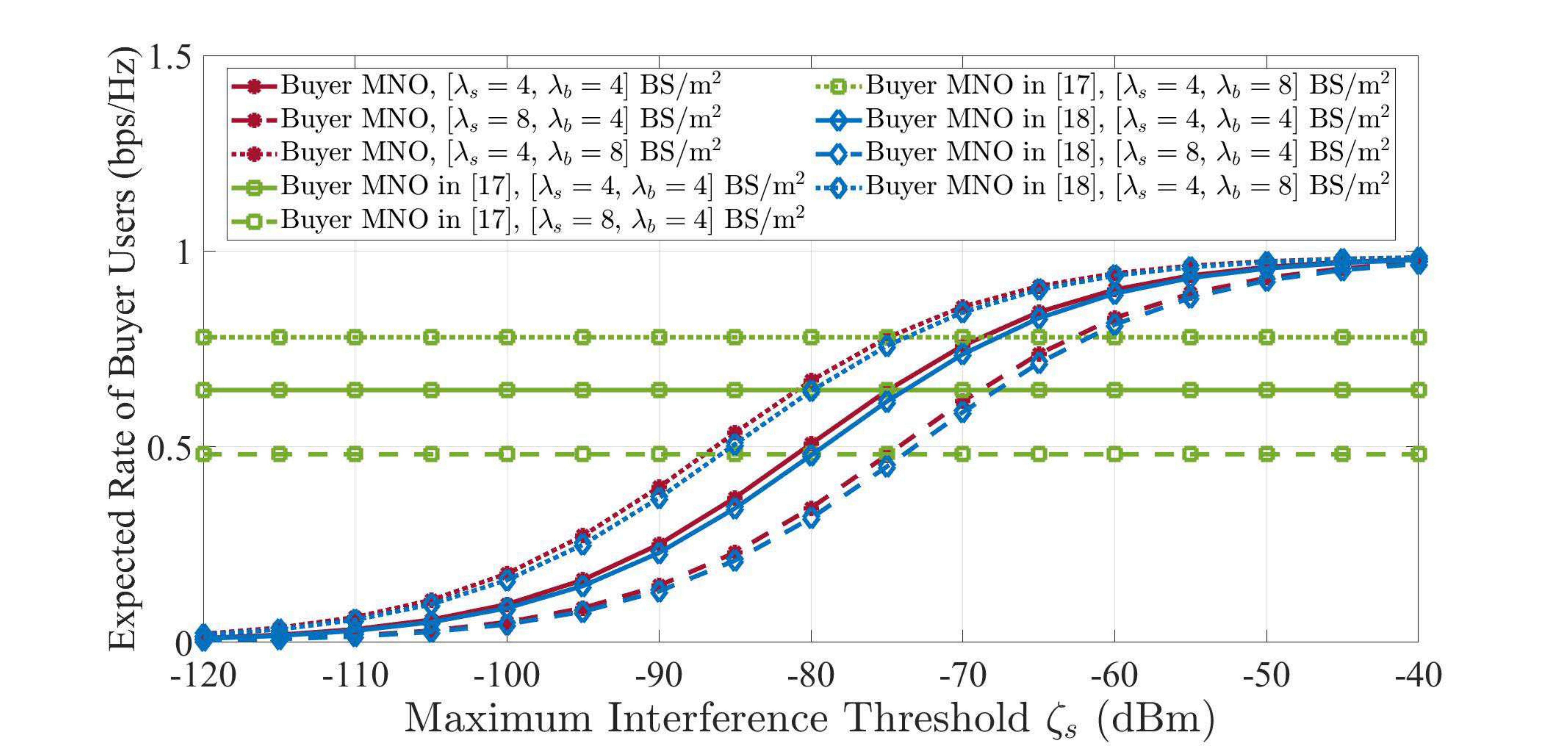}
		\caption{}
		\label{fig:sub-second}
	\end{subfigure}
	\caption{Expected rate of (a) seller users and (b) buyer users versus the maximum interference threshold with different algorithms proposed in \cite{Sanguanpuak2017Conf} and \cite{Gupta2016}.}
	\label{demo9}
	\end{figure}
	\begin{figure}[!t]
		\centering
		\includegraphics[width=9.5 cm,height=6 cm]{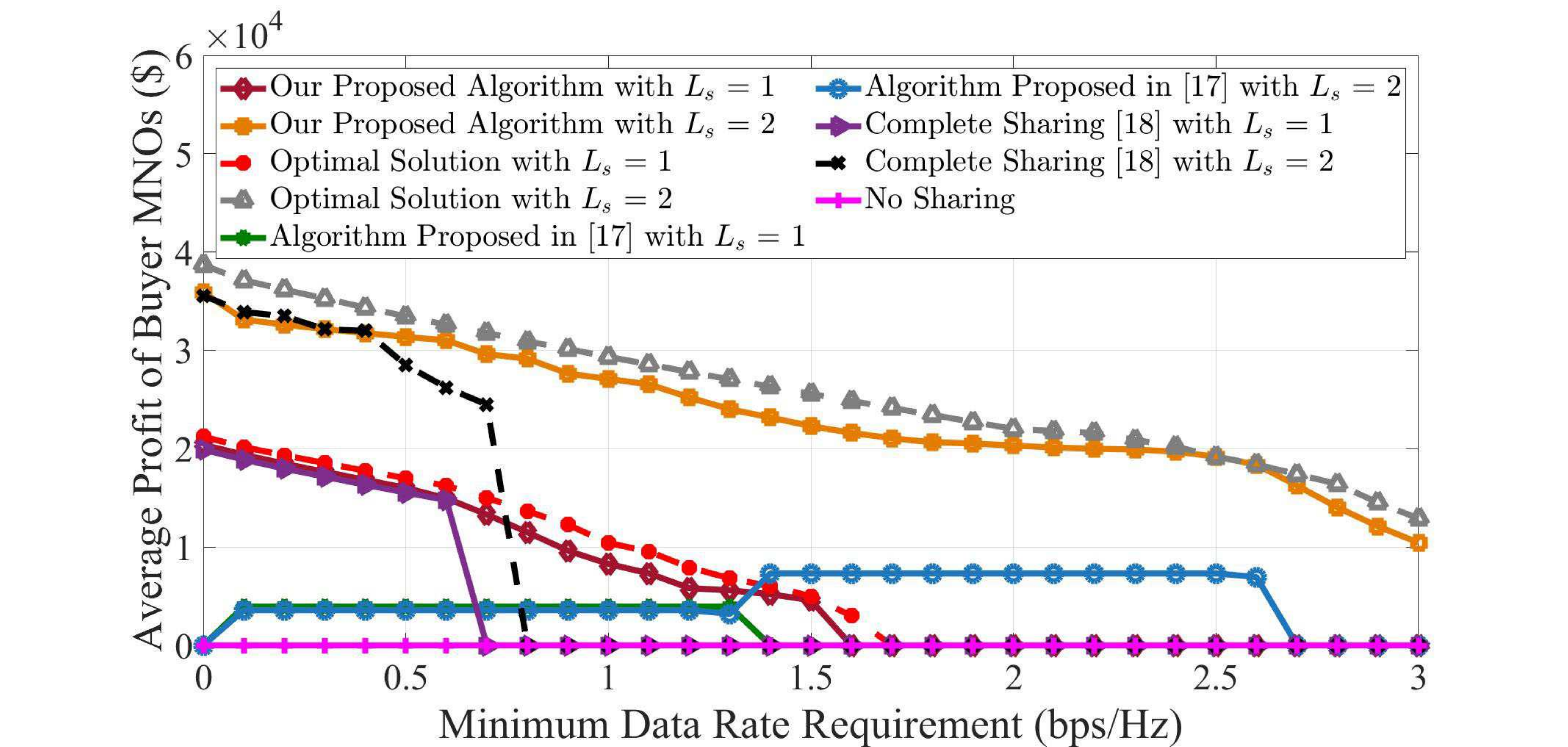}
		\caption{Comparison of the proposed algorithm with the algorithms proposed in \cite{Sanguanpuak2017Conf} and \cite{Gupta2016}. \label{demo8}}
	\end{figure}
	\subsection{Comparing the Performance of the Proposed Algorithm with Existing Algorithms}
	In Fig. \ref{demo9}, we compare the performance of the proposed algorithm with the algorithms presented in \cite{Sanguanpuak2017Conf} and \cite{Gupta2016} versus the maximum interference threshold $\zeta_{s}$ for different values of BS intensity. For generating this figure, we consider a single seller and buyer MNO where the seller MNO has one licensed sub-band, as in \cite{Sanguanpuak2017Conf} and \cite{Gupta2016}. As observed in Fig. \ref{fig:sub-first} and Fig. \ref{fig:sub-second}, the proposed algorithm outperforms the other algorithms presented in \cite{Sanguanpuak2017Conf} and \cite{Gupta2016}. In fact, the proposed algorithm adjusts the transmit power of buyer MNO's BSs in such a way that the interference imposed on all corresponding seller users is below $\zeta_{s}$. Thus, the expected rate of seller users improves significantly. Furthermore, in Fig. \ref{fig:sub-first} and Fig. \ref{fig:sub-second}, we can see that the expected rates of buyer and seller users remain constant in \cite{Sanguanpuak2017Conf} when the maximum interference threshold $\zeta_{s}$ increases. The reason for this is that there is no power control strategy in \cite{Sanguanpuak2017Conf}.
		
	In Fig. \ref{demo8}, we compare the performance of the proposed algorithm with the algorithm presented in \cite{Sanguanpuak2017Conf}, the complete sharing algorithm in \cite{Gupta2016}, and no sharing algorithm versus the minimum data rate requirements ($r^{\text{th}}$) for different numbers of licensed sub-bands per seller MNO ($L_{s}, s\in\mathcal{S}$), as in \cite{Sanguanpuak2017Conf}, \cite{Gupta2016} and \cite{Sanguanpuak2017}. For a fair comparison, we consider a cellular network in which there is a single buyer MNO. In addition, we omit constraints C0 and C2 in problem (\ref{Problem_5}). For generating this figure, we set the number of seller MNOs to 2 ($S = 2$), where the seller MNOs may lease all of their licensed sub-bands to the buyer MNO. As observed in Fig. \ref{demo8}, the proposed algorithm outperforms the other algorithms. The reason is that the algorithm presented in \cite{Sanguanpuak2017Conf} is a greedy algorithm that minimizes the total cost of purchasing licensed sub-bands (i.e., $\min_{\boldsymbol{\rm{A}}}c_{b}(\boldsymbol{\rm{A}}), \forall b\in\mathcal{B}$, where $|\mathcal{B}|=1$), while the proposed algorithm maximizes the profit of the buyer MNO by a joint power control and licensed sub-band sharing. From Fig. \ref{demo8}, we can see that the average profit of the buyer MNOs is zero with no sharing, while sharing all sub-bands among MNOs does not necessarily increase efficiency. Another important observation from Fig. \ref{demo8} is that the stationary point obtained by the proposed algorithm is considerably close to the globally optimal solution.

	\section{Conclusion}\label{Con}
	In this paper, we studied the nonorthogonal spectrum sharing among multiple seller and buyer MNOs, where each seller MNO may lease multiple licensed sub-bands to buyer MNOs. To achieve this, we first analyzed the expected rate of a typical user for each seller and buyer MNO by employing stochastic geometry. Then, we expressed the expected profit function of each MNO in terms of its users' expected rate and the price of leased sub-bands. Next, we formally stated the joint problem of power control and licensed sub-band sharing for maximizing the profit of all MNOs as a MOOP when the QoS constraint of users and the nonnegative return on investment of buyer MNOs were satisfied. In the original form, the stated problem was a MOOP and nonconvex. To tackle these difficulties, we first transformed it into a problem with a single objective function. Then, we relaxed the binary sub-band sharing variables by adding a penalty function to the objective function. Next, by applying the CSSCA algorithm, we proposed an iterative algorithm to address the problem in which, at each iteration, a convex problem using convex surrogate functions was solved. Finally, we illustrated the convergence and performance of the proposed algorithm with numerical results.

\begin{IEEEbiography}[{\includegraphics[width=1in,height=1.25in,clip,keepaspectratio]{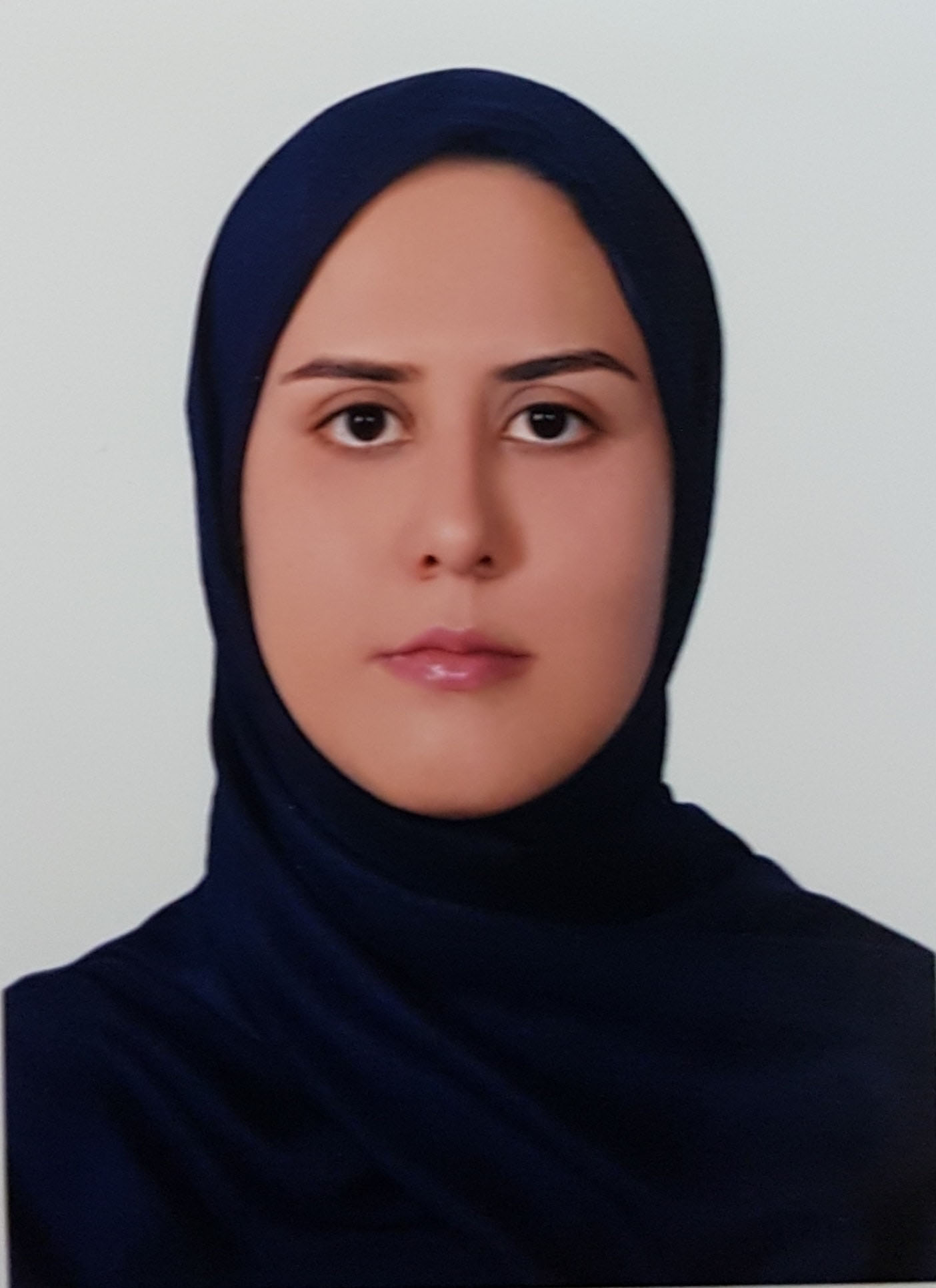}}]{Elaheh Ataeebojd} received her B.Sc. degree from University of Sistan and Baluchestan, Zahedan, Iran, and her M.Sc. degree from University of Isfahan in 2010 and 2016, respectively. She is currently pursuing her Ph.D. degree in Amirkabir University of Technology, Tehran, Iran. Her current research interests include resource management in wireless networks, stochastic geometry, and optimization.
\end{IEEEbiography}

\begin{IEEEbiography}[{\includegraphics[width=1in,height=1.25in,clip,keepaspectratio]{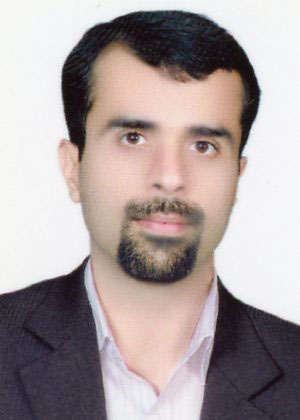}}]{Mehdi Rasti} (S'08-M'11-SM'21) is currently an Associated Professor at the Department of Computer Engineering, Amirkabir University of Technology, Tehran, Iran and is a visiting researcher at the Lappeenranta-Lahti University of Technology (LUT), Lappeenranta, Finland. From November 2007 to November 2008, he was a visiting researcher at the Wireless@KTH, Royal Institute of Technology, Stockholm, Sweden. From September 2010 to July 2012 he was with Shiraz University of Technology, Shiraz, Iran. From June 2013 to August 2013, and from July 2014 to August 2014 he was a visiting researcher in the Department of Electrical and Computer Engineering, University of Manitoba, Winnipeg, MB, Canada. He received his B.Sc. degree from Shiraz University, Shiraz, Iran, and the M.Sc. and Ph.D. degrees both from Tarbiat Modares University, Tehran, Iran, all in Electrical Engineering in 2001, 2003 and 2009, respectively. His current research interests include radio resource allocation in IoT, Beyond 5G and 6G wireless networks.
\end{IEEEbiography}

\begin{IEEEbiography}[{\includegraphics[width=1in,height=1.25in,clip,keepaspectratio]{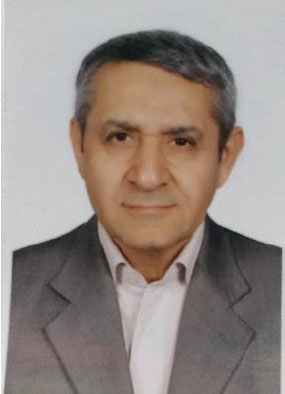}}]{Hossein Pedram}
	Associate professor of Computer Engineering (retired) at
	Amirkabir University of Technology. He received his B.Sc. degree in Electrical Engineering from Sharif University in 1979, and received M.Sc. and Ph.D. degrees from Ohio State University and Washington State University respectively in Computer Engineering. Dr. Pedram served as a faculty member in the Computer Engineering	Department at Amirkabir University of Technology from 1992 to 2018. He teaches a
	variety of courses in Computer Engineering including Computer architecture, Computer Networks, Operating Systems and distributed systems. His research interests include data communications, innovations in computer architecture, nanoscale digital circuits, management of computer networks, and distributed systems. Dr. Pedram can be reached through hpedram@uw.edu.
\end{IEEEbiography}

\begin{IEEEbiography}[{\includegraphics[width=1in,height=1.25in,clip,keepaspectratio]{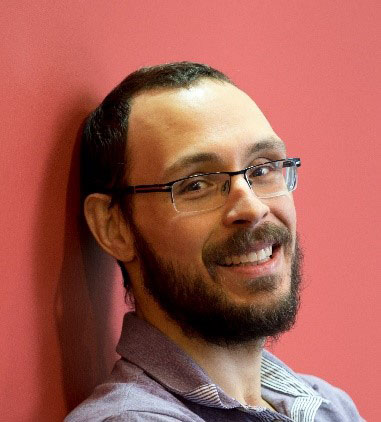}}]{Pedro H. J. Nardelli} [M'07, SM'19] received the B.S. and M.Sc. degrees in electrical engineering from the State University of Campinas, Brazil, in 2006 and 2008, respectively. In 2013, he received his doctoral degree from University of Oulu, Finland, and State University of Campinas following a dual degree agreement. He is currently Associate Professor (tenure track) in IoT in Energy Systems at LUT University, Finland, and holds a position of Academy of Finland Research Fellow with a project called Building the Energy Internet as a large-scale IoT-based cyber-physical system that manages the energy inventory of distribution grids as discretized packets via machine-type communications (EnergyNet). He leads the \href{https://cps-g.com/}{Cyber-Physical Systems Group}  at LUT, and is Project Coordinator of the CHIST-ERA European consortium Framework for the Identification of Rare Events via Machine Learning and IoT Networks (FIREMAN) and of the project Swarming Technology for Reliable and Energy-aware Aerial Missions (STREAM) supported by Jane and Aatos Erkko Foundation. He is also Docent at University of Oulu in the topic of \enquote{communications strategies and information processing in energy systems}. His research focuses on wireless communications particularly applied in industrial automation and energy systems. He received a best paper award of IEEE PES Innovative Smart Grid Technologies Latin America 2019 in the track \enquote{Big Data and Internet of Things}. He is also IEEE Senior Member. More information: \url{https://sites.google.com/view/nardelli/}
\end{IEEEbiography}

\end{document}